# The basic physics of the binary black hole merger GW150914

*LIGO Scientific and VIRGO Collaborations*[*,**]

The first direct gravitational-wave detection was made by the Advanced Laser Interferometer Gravitational Wave Observatory on September 14, 2015. The GW150914 signal was strong enough to be apparent, without using any waveform model, in the filtered detector strain data. Here, features of the signal visible in the data are analyzed using concepts from Newtonian physics and general relativity, accessible to anyone with a general physics background. The simple analysis presented here is consistent with the fully general-relativistic analyses published elsewhere, in showing that the signal was produced by the inspiral and subsequent merger of two black holes. The black holes were each of approximately $35\,\mathrm{M}_\odot$, still orbited each other as close as $\sim 350$ km apart and subsequently merged to form a single black hole. Similar reasoning, directly from the data, is used to roughly estimate how far these black holes were from the Earth, and the energy that they radiated in gravitational waves.

## 1 Introduction

Advanced LIGO made the first observation of a gravitational wave (GW) signal, GW150914 [1], on September 14th, 2015, a successful confirmation of a prediction by Einstein's theory of general relativity (GR). The signal was clearly seen by the two LIGO detectors located in Hanford, WA and Livingston, LA. Extracting the full information about the source of the signal requires detailed analytical and computational methods (see [2–6] and references therein for details). However, much can be learned about the source by direct inspection of the detector data and some basic physics [7], accessible to a general physics audience, as well as students and teachers. This simple analysis indicates that the source is two black holes (BHs) orbiting around one another and then merging to form another black hole.

A black hole is a region of space-time where the gravitational field is so intense that neither matter nor radiation can escape. There is a natural "gravitational radius" associated with a mass $m$, called the Schwarzschild radius, given by

$$r_{\mathrm{Schwarz}}(m) = \frac{2Gm}{c^2} = 2.95\left(\frac{m}{\mathrm{M}_\odot}\right)\,\mathrm{km}, \tag{1}$$

where $\mathrm{M}_\odot = 1.99 \times 10^{30}$kg is the mass of the Sun, $G = 6.67 \times 10^{-11}\ \mathrm{m}^3/\mathrm{s}^2\mathrm{kg}$ is Newton's gravitational constant, and $c = 2.998 \times 10^8$ m/s is the speed of light. According to the hoop conjecture, if a non-spinning mass is compressed to within that radius, then it must form a black hole [8]. Once the black hole is formed, any object that comes within this radius can no longer escape out of it.

Here, the result that GW150914 was emitted by the inspiral and merger of two black holes follows from (1) the strain data visible at the instrument output, (2) dimensional and scaling arguments, (3) primarily Newtonian orbital dynamics and (4) the Einstein quadrupole formula for the luminosity of a gravitational wave source.[1] These calculations are straightforward enough that they can be readily verified with pencil and paper in a short time. Our presentation is by design approximate, emphasizing simple arguments.

Specifically, while the orbital motion of two bodies is approximated by Newtonian dynamics and Kepler's laws to high precision at sufficiently large separations and sufficiently low velocities, we will invoke Newtonian dynamics to describe the motion even toward the end point of orbital motion (We revisit this assumption in

* lvc.publications@ligo.org
** Full author list appears at the end.

1 In the terminology of GR corrections to Newtonian dynamics, (3) & (4) constitute the "0th post-Newtonian" approximation (0PN) (see Sec. 4.4). A similar approximation was used for the first analysis of binary pulsar PSR 1913+16 [9, 10].

Sec. 4.4). The theory of general relativity is a fully non-linear theory, which could make any Newtonian analysis wholly unreliable; however, solutions of Einstein's equations using numerical relativity (NR) [11–13] have shown that a binary system's departures from Newtonian dynamics can be described well using a quantifiable analytic perturbation until quite late in its evolution - late enough for our argument (as shown in Sec. 4.4).

The approach presented here, using basic physics, is intended as a pedagogical introduction to the physics of gravitational wave signals, and as a tool to build intuition using rough, but straightforward, checks. Our presentation here is by design elementary, but gives results consistent with more advanced treatments. The fully rigorous arguments, as well as precise numbers describing the system, have already been published elsewhere [2–6].

The paper is organized as follows: our presentation begins with the data output by the detectors.[2] Section 2 describes the properties of the signal read off the strain data, and how they determine the quantities relevant for analyzing the system as a binary inspiral. We then discuss in Sec. 3, using the simplest assumptions, how the binary constituents must be heavy and small, consistent only with being black holes. In Sec. 4 we examine and justify the assumptions made, and constrain both masses to be well above the heaviest known neutron stars. Section 5 uses the peak gravitational wave luminosity to estimate the distance to the source, and calculates the total luminosity of the system. The appendices provide a calculation of gravitational radiation strain and radiated power (App. A), and discuss astrophysical compact objects of high mass (App. B) as well as what one might learn from the waveform after the peak (App. C).

## 2 Analyzing the observed data

Our starting point is shown in Fig. 1: the instrumentally observed strain data $h(t)$, after applying a band-pass filter to the LIGO sensitive frequency band (35–350 Hz), and a band-reject filter around known instrumental noise frequencies [14]. The time-frequency behavior of the signal is depicted in Fig. 2. An approximate version of the time-frequency evolution can also be obtained directly from the strain data in Fig. 1 by measuring the time differences $\Delta t$ between successive zero-crossings[3] and estimating $f_{\mathrm{GW}} = 1/(2\Delta t)$, without assuming a waveform model. We plot the $-8/3$ power of these estimated frequencies in Fig. 3, and explain its physical relevance below.

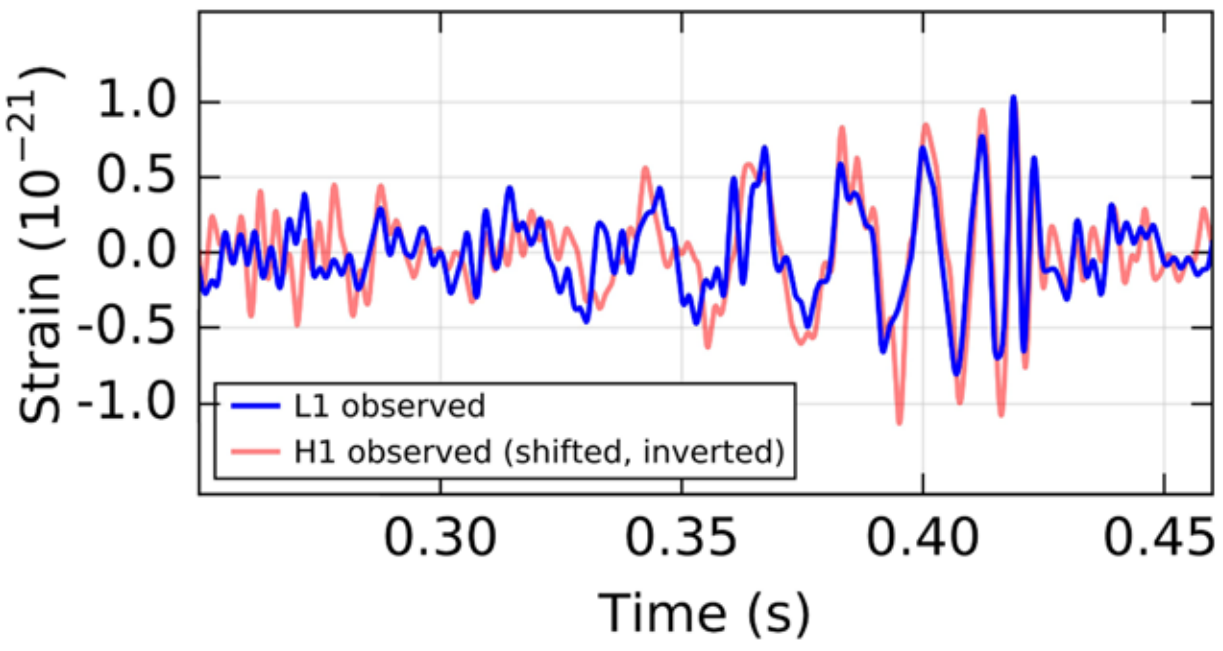

**Figure 1** The instrumental strain data in the Livingston detector (blue) and Hanford detector (red), as shown in Figure 1 of [1]. Both have been bandpass- and notch-filtered. The Hanford strain has been shifted back in time by 6.9 ms and inverted. Times shown are relative to 09:50:45 Coordinated Universal Time (UTC) on September 14, 2015.

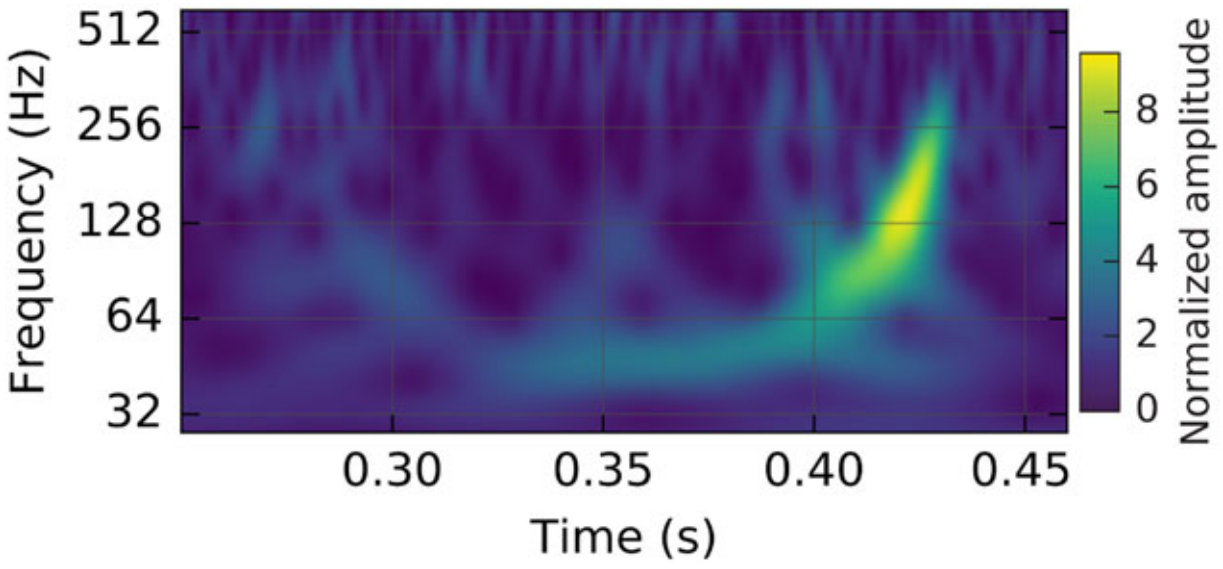

**Figure 2** A representation of the strain-data as a time-frequency plot (taken from [1]), where the increase in signal frequency ("chirp") can be traced over time.

The signal is dominated by several cycles of a wave pattern whose amplitude is initially increasing, starting from around the time mark 0.30 s. In this region the gravitational wave period is decreasing, thus the frequency is increasing. After a time around 0.42 s, the amplitude drops rapidly, and the frequency appears to stabilize. The last clearly visible cycles (in both detectors, after accounting for a 6.9 ms time-of-flight-delay [1]) indicate that the final instantaneous frequency is above 200 Hz. The entire visible part of the signal lasts for around 0.15s.

In general relativity, gravitational waves are produced by accelerating masses [15]. Since the waveform clearly shows at least eight oscillations, we know that a mass or masses are oscillating. The increase in gravitational wave frequency and amplitude also indicate that during this time the oscillation frequency of the source system is increasing. This initial behavior cannot be due to a perturbed system returning back to stable equilibrium, since oscillations around equilibrium are generically characterized by roughly constant frequencies and decaying amplitudes. For example, in the case of a fluid ball, the oscillations would be damped by viscous forces. Here, the data demonstrate very different behavior.

[2] The advanced LIGO detectors use laser interferometry to measure the strain caused by passing gravitational waves. For details of how the detectors work, see [1] and its references.

[3] To resolve the crossing at $t \sim 0.35$ s, when the signal amplitude is lower and the true waveform's sign transitions are difficult to pinpoint, we averaged the positions of the five adjacent zero-crossings (over $\sim 6$ ms).

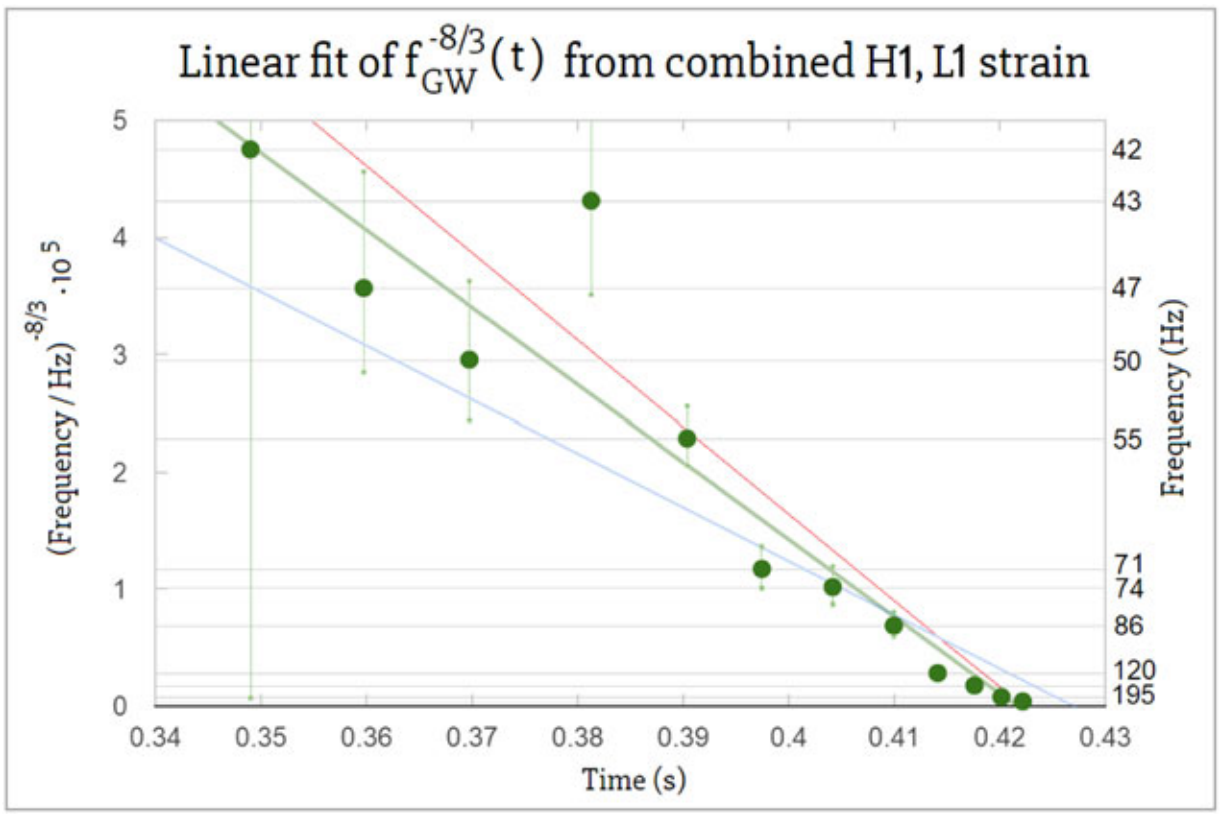

**Figure 3** A linear fit (green) of $f_{\rm GW}^{-8/3}(t)$. While this interpolation used the combined strain data from H1 and L1 (in fact, the sum of L1 with time shifted and sign-flipped H1, as explained). A similar fit can be done using either H1 or L1 strain independently. The fit shown has residual sum of squares $R^2_{\rm L1-H1} \sim 0.9$; we have also found $R^2_{\rm H1} \sim 0.9$ and $R^2_{\rm L1} \sim 0.8$. The slope of this fitted line gives an estimate of the chirp mass of $\sim 37\,{\rm M}_\odot$ using Eq. 8. The blue and red lines indicate $\mathscr{M}$ of $30{\rm M}_\odot$ and $40{\rm M}_\odot$, respectively. The error-bars have been estimated by repeating the procedure for waves of the same amplitudes and frequencies added to the LIGO strain data just before GW150914. A similar error estimate has been found using the differences between H1 and L1 zero-crossings.

During the period when the gravitational wave frequency and amplitude are increasing, orbital motion of two bodies is the only plausible explanation: there, the only "damping forces" are provided by gravitational wave emission, which brings the orbiting bodies closer (an "inspiral"), increasing the orbital frequency and amplifying the gravitational wave energy output from the system.[4]

Gravitational radiation has many aspects analogous to electromagnetic (EM) radiation from accelerating charges. A significant difference is that there is no analog to EM dipole radiation, whose amplitude is proportional to the second time derivative of the electric dipole moment. This is because the gravitational analog is the mass dipole moment ($\sum_A m_A \mathbf{x}_A$ at leading order in the velocity) whose first time derivative is the total linear momentum, which is conserved for a closed system, and whose second derivative therefore vanishes. Hence, at leading order, gravitational radiation is quadrupolar. Because the quadrupole moment (defined in App. A) is symmetric under rotations by $\pi$ about the orbital axis, the radiation has a frequency *twice* that of the orbital frequency (for a detailed calculation for a 2-body system, see App. A and pp. 356-357 of [16]).

The eight gravitational wave cycles of increasing frequency therefore require at least four orbital revolutions, at separations large enough (compared to the size of the bodies) that the bodies do not collide. The rising frequency signal eventually terminates, suggesting the end of inspiraling orbital motion. As the amplitude decreases and the frequency stabilizes the system returns to a stable equilibrium configuration. We shall show that the only reasonable explanation for the observed frequency evolution is that the system consisted of two black holes that orbited each other and subsequently merged.

**Determining the frequency at maximum strain amplitude** $f_{\rm GW}|_{\rm max}$: The single most important quantity for the reasoning in this paper is the gravitational wave frequency at which the waveform has maximum amplitude. Using the zero-crossings around the peak of Fig. 1 and/or the brightest point of Fig. 2, we take the conservative (low) value

$$f_{\rm GW}\big|_{\rm max} \sim 150\ {\rm Hz}, \tag{2}$$

where here and elsewhere the notation indicates that the quantity before the vertical line is evaluated at the time indicated after the line. We thus interpret the observational data as indicating that the bodies were orbiting each other (roughly Keplerian dynamics) up to at least an orbital angular frequency

$$\omega_{\rm Kep}\big|_{\rm max} = \frac{2\pi\, f_{\rm GW}\big|_{\rm max}}{2} = 2\pi\times 75\ {\rm Hz}. \tag{3}$$

**Determining the mass scale**: Einstein found [17] that the gravitational wave strain $h$ at a (luminosity) distance $d_L$ from a system whose traceless mass quadrupole moment is $Q_{ij}$ (defined in App. A) is

$$h_{ij} = \frac{2\,G}{c^4\,d_L}\frac{{\rm d}^2\,Q_{ij}}{{\rm d}t^2}, \tag{4}$$

and that the rate at which energy is carried away by these gravitational waves is given by the quadrupole

[4] The possibility of a different inspiraling system, whose evolution is not governed by gravitational waves, is explored in App. A.1 and shown to be inconsistent with this data.

formula [17]

$$\frac{\mathrm{d}E_{\mathrm{GW}}}{\mathrm{d}t} = \frac{c^3}{16\pi\, G}\iint |\dot{h}|^2 \mathrm{d}S = \frac{1}{5}\frac{G}{c^5}\sum_{i,\,j=1}^{3}\frac{\mathrm{d}^3 Q_{ij}}{\mathrm{d}t^3}\frac{\mathrm{d}^3 Q_{ij}}{\mathrm{d}t^3}, \quad (5)$$

$$\text{where } |\dot{h}|^2 = \sum_{i,\,j=1}^{3}\frac{\mathrm{d}h_{ij}}{\mathrm{d}t}\frac{\mathrm{d}h_{ij}}{\mathrm{d}t},$$

the integral is over a sphere at radius $d_L$ (contributing a factor $4\pi d_L^2$), and the quantity on the right-hand side must be averaged over (say) one orbit.[5]

In our case, Eq. 5 gives the rate of loss of orbital energy to gravitational waves, when the velocities of the orbiting objects are not too close to the speed of light, and the strain is not too large [15]; we will apply it until the frequency $f_{\mathrm{GW}}|_{\mathrm{max}}$, see Sec. 4.4. This wave description is applicable in the "wave zone" [19], where the gravitational field is weak and the expansion of the universe is ignored (see Sec. 4.6).

For the binary system we denote the two masses by $m_1$ and $m_2$, the total mass by $M = m_1 + m_2$, and the reduced mass by $\mu = m_1 m_2/M$. We define the mass ratio $q = m_1/m_2$ and without loss of generality assume that $m_1 \geq m_2$ so that $q \geq 1$. To describe the gravitational wave emission from a binary system, a useful mass quantity is the *chirp mass*, $\mathcal{M}$, related to the component masses by

$$\mathcal{M} = \frac{(m_1 m_2)^{3/5}}{(m_1 + m_2)^{1/5}}. \quad (6)$$

Using Newton's laws of motion, Newton's universal law of gravitation, and Einstein's quadrupole formula for the gravitational wave luminosity of a system, a simple formula is derived in App. A (following [20, 21]) relating the frequency and frequency derivative of emitted gravitational waves to the chirp mass,

$$\mathcal{M} = \frac{c^3}{G}\left(\left(\frac{5}{96}\right)^3 \pi^{-8}\,(f_{\mathrm{GW}})^{-11}\left(\dot{f}_{\mathrm{GW}}\right)^3\right)^{1/5}, \quad (7)$$

where $\dot{f}_{\mathrm{GW}} = \mathrm{d}f_{\mathrm{GW}}/\mathrm{d}t$ is the rate-of-change of the frequency (see Eq. A5 and Eq. 3 of [22]). This equation is expected to hold as long as the Newtonian approximation is valid (see Sec. 4.4).

Thus, a value for the chirp mass can be determined directly from the observational data, using the frequency and frequency derivative of the gravitational waves at any moment in time. For example, values for the frequency can be estimated from the time-frequency plot of the observed gravitational wave strain data (Fig. 2), and for the frequency derivative by drawing tangents to the same curve (see figure on journal cover). The time interval during which the inspiral signal is in the sensitive band of the detector (and hence is visible) corresponds to gravitational wave frequencies in the range $30 < f_{\mathrm{GW}} < 150$ Hz. Over this time, the frequency (period) varies by a factor of 5 ($\frac{1}{5}$), and the frequency derivative varies by more than two orders-of-magnitude. The implied chirp mass value, however, remains constant to within 35%. The exact value of $\mathcal{M}$ is not critical to the arguments that we present here, so for simplicity we take $\mathcal{M} = 30\,\mathrm{M}_\odot$.

**Note that the characteristic mass scale of the radiating system is obtained by direct inspection of the time-frequency behavior of the observational data.**

The fact that the chirp mass remains approximately constant for $f_{\mathrm{GW}} < 150\mathrm{Hz}$ is strong support for the orbital interpretation. The fact that the amplitude of the gravitational wave strain increases with frequency also supports this interpretation, and suggests that the assumptions that go into the calculation which leads to these formulae are applicable: the velocities in the binary system are not too close to the speed of light, and the orbital motion has an adiabatically changing radius and a period described instantaneously by Kepler's laws. The data also indicate that these assumptions certainly break down at a gravitational wave frequency above $f_{\mathrm{GW}}|_{\mathrm{max}}$, as the amplitude stops growing.

Alternatively, Eq. 7 can be integrated to obtain

$$f_{\mathrm{GW}}^{-8/3}(t) = \frac{(8\pi)^{8/3}}{5}\left(\frac{G\mathcal{M}}{c^3}\right)^{5/3}(t_c - t), \quad (8)$$

which does not involve $\dot{f}_{\mathrm{GW}}$ explicitly, and can therefore be used to calculate $\mathcal{M}$ directly from the time periods between zero-crossings in the strain data. The constant of integration $t_c$ is the time of coalescense. We have performed such an analysis, presented in Fig. 3, to find similar results. We henceforth adopt a conservative lower estimate of $\mathcal{M} = 30\,\mathrm{M}_\odot$ for the chirp mass. We remark that this mass is derived from quantities measured in the detector frame, thus it and the quantities we derive from it are given in the detector frame. Discussion of redshift from the source frame appears in Sec. 4.6.

[5] See App. A for a worked-out calculation, and pp. 974-977 of [18] for a derivation of these results, obtained by linearizing the Einstein Equation, the central equation of general relativity.

## 3 Evidence for compactness in the simplest case

For simplicity, suppose that the two bodies have equal masses, $m_1 = m_2$. The value of the chirp mass then implies that $m_1 = m_2 = 2^{1/5}\mathscr{M} = 35\,\mathrm{M}_\odot$, so that the total mass would be $M = m_1 + m_2 = 70\,\mathrm{M}_\odot$. We also assume for now that the objects are not spinning, and that their orbits remain Keplerian and essentially circular until the point of peak amplitude.

Around the time of peak amplitude the bodies therefore had an orbital separation $R$ given by

$$R = \left(\frac{GM}{\omega^2_{\mathrm{Kep}}\big|_{\max}}\right)^{1/3} = 350\,\mathrm{km}. \tag{9}$$

Compared to normal length scales for stars, this is a *tiny* value. This constrains the objects to be exceedingly small, or else they would have collided and merged long before reaching such close proximity. Main-sequence stars have radii measured in hundreds of thousands or millions of kilometers, and white dwarf (WD) stars have radii which are typically ten thousand kilometers. Scaling Eq. 9 shows that such stars' inspiral evolution would have terminated with a collision at an orbital frequency of a few mHz (far below 1 Hz).

The most compact stars known are neutron stars, which have radii of about ten kilometers. Two neutron stars could have orbited at this separation without colliding or merging together – but the maximum mass that a neutron star can have before collapsing into a black hole is about $3\,\mathrm{M}_\odot$ (see App. B).

In our case, the bodies of mass $m_1 = m_2 = 35\,\mathrm{M}_\odot$ each have a Schwarzschild radius of 103 km. This is illustrated in Fig. 4. The orbital separation of these objects, 350 km, is only about twice the sum of their Schwarzschild radii.

In order to quantify the closeness of the two objects relative to their natural gravitational radius, we introduce the compactness ratio $\mathscr{R}$. This is defined as the Newtonian orbital separation between the centers of the objects divided by the sum of their smallest possible respective radii (as compact objects). For the non-spinning, circular orbit, equal-mass case just discussed $\mathscr{R} = 350\,\mathrm{km}/206\,\mathrm{km} \sim 1.7$.

For comparison with other known Keplerian systems, the orbit of Mercury, the innermost planet in our solar system, has $\mathscr{R} \sim 2 \times 10^7$, the binary orbit for the stellar black hole in Cyg X-1[6] has $\mathscr{R} \sim 3 \times 10^5$, and the binary system of highest known orbital frequency, the WD system HM Cancri (RX J0806), has $\mathscr{R} \sim 2 \times 10^4$ [24]. Observations of orbits around our galactic center indicate the presence of a supermassive black hole, named Sgr A* [25, 26], with the star S2 orbiting it as close as $\mathscr{R} \sim 10^3$. For a system of two neutron stars just touching, $\mathscr{R}$ would be between $\sim 2$ and $\sim 5$.

The fact that the Newtonian/Keplerian evolution of the orbit inferred from the signal of GW150914 breaks down when the separation is about the order of the black hole radii (compactness ratio $\mathscr{R}$ of order 1) is further evidence that the objects are highly compact.

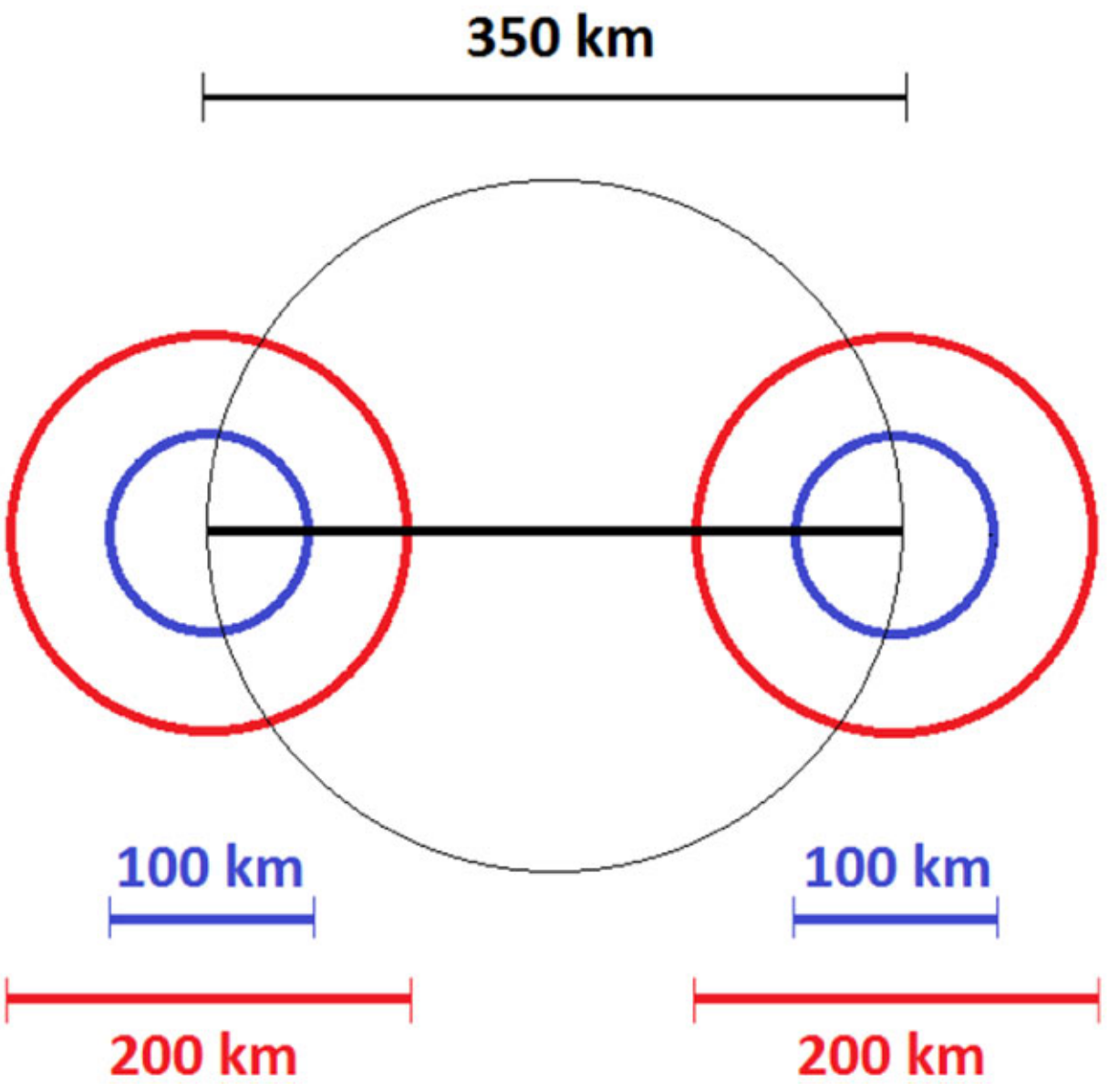

**Figure 4** A demonstration of the scale of the orbit at minimal separation (black, 350 km) vs. the scale of the compact radii: Schwarzschild (red, diameter 200 km) and extremal Kerr (blue, diameter 100 km). Note the masses here are equal; as Sec. 4.2 explains, the system is even more compact for unequal masses. While identification of a rigid reference frame for measuring distances between points is not unique in relativity, this complication only really arises with strong gravitational fields, while in the Keplerian regime (of low compactness and low gravitational potentials) the system's center-of-mass rest-frame can be used. Therefore if the system is claimed to be non-compact, the Keplerian argument should hold, and constrain the distances to be compact. Thus the possibility of non-compactness is inconsistent with the data; see also Sec. 4.4.

6 Radio, optical and X-ray telescopes have probed the accretion disk extending much further inside [23].

Original Paper

## 4 Revisiting the assumptions

In Sec. 3 we used the data to show that the coalescing objects are black holes under the assumptions of a circular orbit, equal masses, and no spin. It is not possible, working at the level of approximation that we are using here, to directly constrain these parameters of the system (although more advanced techniques are able to constrain them, see [2]). However, it is possible to examine how these assumptions affect our conclusions and in this section we show that relaxing them does not significantly change the outcome. We also use the Keplerian approximation to discuss these three modifications (Sec. 4.1–4.3), then revisit the Keplerian assumption itself, and discuss the consequences of foregoing it (Sec. 4.4–4.5). In Sec. 4.6 we discuss the distance to the source, and its potential effects.

### 4.1 Orbital eccentricity

For non-circular orbits with eccentricity $e > 0$, the $R$ of Kepler's third law (Eq. 9) no longer refers to the orbital separation but rather to the semi-major axis. The instantaneous orbital separation $r_{\rm sep}$ is bounded from above by $R$, and from below by the point of closest approach (periapsis), $r_{\rm sep} \geq (1-e)R$. We thus see that the compactness bound imposed by eccentric orbits is even tighter (the compactness ratio $\mathscr{R}$ is smaller).

There is also a correction to the luminosity which depends on the eccentricity. However, this correction is significant only for highly eccentric orbits.[7] For these, the signal should display a modulation [27]: the velocity would be greater near periapsis than near apoapsis, so the signal would alternate between high-amplitude and low-amplitude peaks. Such modulation is not seen in the data, whose amplitude grows monotonically.

This is not surprising, as the angular momentum that gravitational waves carry away causes the orbits to circularize much faster than they shrink [20, 21]. This correction can thus be neglected.

[7] Eccentricity increases the luminosity [20, 21] by a factor $\ell(e) = \left(1-e^2\right)^{-7/2}\left(1+\frac{73}{24}e^2+\frac{37}{96}e^4\right) \geq 1$, thus reducing the chirp mass (inferred using Eq. 7) to $\mathscr{M}(e) = \ell^{-3/5}(e)\cdot\mathscr{M}(e{=}0)$. Taking into account the ratio between the separation at periapsis and the semi-major axis, one obtains $\mathscr{R}(e) = \left(1-e\right)\ell^{2/5}(e)\cdot\mathscr{R}(e{=}0)$. Hence for the compactness ratio to increase, the eccentricity must be $e \gtrsim 0.6$, and for a factor of 2, $e \gtrsim 0.9$ (see Fig. 5)

### 4.2 The case of unequal masses

It is easy to see that the compactness ratio $\mathscr{R}$ also gets smaller with increasing mass-ratio, as that implies a higher total mass for the observed value of the Newtonian order chirp mass. To see this explicitly, we express the component masses and total mass in terms of the chirp mass $\mathscr{M}$ and the mass ratio $q$, as $m_1 = \mathscr{M}(1+q)^{1/5}q^{2/5}$, $m_2 = \mathscr{M}(1+q)^{1/5}q^{-3/5}$, and

$$M = m_1 + m_2 = \mathscr{M}(1+q)^{6/5}q^{-3/5}. \tag{10}$$

The compactness ratio $\mathscr{R}$ is the ratio of the orbital separation $R$ to the sum of the Schwarzschild radii of the two component masses, $r_{\rm Schwarz}(M) = r_{\rm Schwarz}(m_1) + r_{\rm Schwarz}(m_2)$, giving

$$\begin{aligned}\mathscr{R} &= \frac{R}{r_{\rm Schwarz}(M)} = \frac{c^2}{2(\omega_{\rm Kep}|_{\rm max}GM)^{2/3}} \\ &= \frac{c^2}{2(\pi f_{\rm GW}|_{\rm max}G\mathscr{M})^{2/3}}\frac{q^{2/5}}{(1+q)^{4/5}} \approx \frac{3.0\,q^{2/5}}{(1+q)^{4/5}}\,. \end{aligned}\tag{11}$$

This quantity is plotted in Fig. 5, which clearly shows that for mass ratios $q > 1$ the compactness ratio *decreases*: the separation between the objects becomes smaller when measured in units of the sum of their Schwarzschild radii. Thus, for a given chirp mass and orbital frequency, a system composed of unequal masses is *more* compact than one composed of equal masses.

One can also place an *upper* limit on the mass ratio $q$, thus a lower bound on the smaller mass $m_2$, based purely on the data. This bound arises from minimal compactness: we see from the compactness ratio plot in Fig. 5 that beyond the mass ratio of $q \sim 13$ the system becomes so compact that it will be within the Schwarzschild radii of the combined mass of the two bodies. This gives us a limit for the mass of the smaller object $m_2 \geq 11\,{\rm M}_\odot$. As this is 3–4 times more massive than the neutron star limit, both bodies are expected to be black holes .

### 4.3 The effect of objects' spins

The third assumption we relax concerns the spins of the objects. For a mass $m$ with spin angular momentum $S$ we define the dimensionless spin parameter

$$\chi = \frac{c}{G}\frac{S}{m^2}. \tag{12}$$

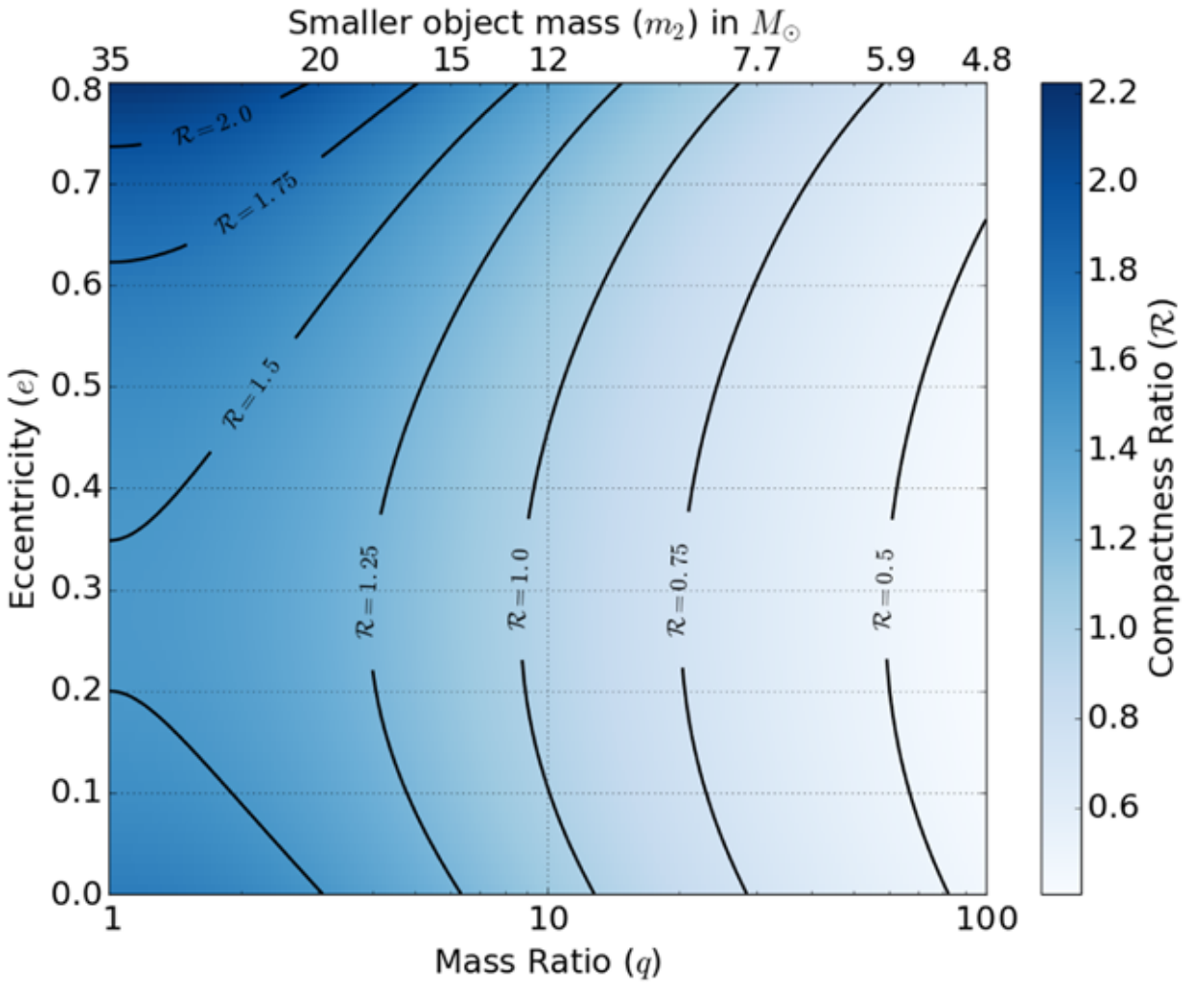

**Figure 5** This figure shows the compactness ratio constraints imposed on the binary system by $\mathscr{M} = 30\,\mathrm{M}_\odot$ and $f_{\rm GW}|_{\rm max} = 150\,\mathrm{Hz}$. It plots the compactness ratio (the ratio of the separation between the two objects to the sum of their Schwarzschild radii) as a function of mass ratio and eccentricity from $e = 0$ to the very high (arbitrary) value of $e = 0.8$. The bottom-left corner ($q = 1$, $e = 0$) corresponds to the case given in Sec. 3. At fixed mass ratio, the system becomes more compact with growing eccentricity until $e = 0.27$, as explained in Sec. 4.1. The bottom edge ($e = 0$) illustrates the argument given in Sec. 4.2 and Eq. 11: the system becomes more compact as the mass ratio increases. We note that (for $e = 0$) beyond mass ratio of $q \sim 13$ ($m_2 \sim 11\,\mathrm{M}_\odot$) the system would become more compact than the sum of the component Schwarzschild radii.

The spins of $m_1$ and $m_2$ modify their gravitational radii as described in this subsection, as well as the orbital dynamics, as described in the next subsection.

The smallest radius a non-spinning object ($\chi = 0$) could have without being a black hole is its Schwarzschild radius. Allowing the objects to have angular momentum (spin) pushes the limit down by a factor of two, to the radius of an extremal Kerr black hole (for which $\chi = 1$), $r_{\rm EK}(m) = \frac{1}{2} r_{\rm Schwarz}(m) = Gm/c^2$. As this is linear in the mass, and summing radii linearly, we obtain a lower limit on the Newtonian separation of two adjacent non-black hole bodies of total mass $M$ is

$$r_{\rm EK}(m_1) + r_{\rm EK}(m_2) = \frac{1}{2} r_{\rm Schwarz}(M) = \frac{G\,M}{c^2} \approx 1.5 \left(\frac{M}{\mathrm{M}_\odot}\right) \mathrm{km}. \tag{13}$$

The compactness ratio can also be defined in relation to $r_{\rm EK}$ rather than $r_{\rm Schwarz}$, which is at most a factor of two larger than for non-spinning objects.

We may thus constrain the orbital compactness ratio (now accounting for eccentricity, unequal masses, and spin) by

$$\begin{aligned} \mathscr{R} &= \frac{r_{\rm sep}(M)}{r_{\rm EK}(M)} \leq \frac{R(M)}{r_{\rm EK}(M)} = \frac{c^2}{\left(G\,M\omega_{\rm Kep}\right)^{2/3}} \\ &\leq \frac{c^2}{\left(2^{6/5}\,G\,\mathscr{M}\,\omega_{\rm Kep}\right)^{2/3}} \\ &= \frac{c^2}{\left(2^{6/5}\,\pi\,G\,\mathscr{M}\,f_{\rm GW}|_{\rm max}\right)^{2/3}} \simeq 3.4, \end{aligned} \tag{14}$$

where in the last step we used $\mathscr{M} = 30\,\mathrm{M}_\odot$ and $f_{\rm GW}|_{\rm max} = 150$ Hz. This constrains the constituents to be under 3.4 (1.7) times their extremal Kerr (Schwarzschild) radii, making them highly compact. The compact arrangement is illustrated in Fig. 4.

We can also derive an upper limit on the value of the mass ratio $q$, from the constraint that the compactness ratio must be larger than unity. This is because, for a fixed value of the chirp mass $\mathscr{M}$ and a fixed value of $f_{\rm GW}|_{\rm max}$, the compactness ratio $\mathscr{R}$ decreases as the mass ratio $q$ increases. Thus, the constraint $\mathscr{R} \geq 1$, puts a limit on the maximal possible $q$ and thus on the maximum total mass $M_{\rm max}$,

$$\left(\frac{M_{\rm max}}{\mathscr{M}}\right) \simeq 3.4^{3/2} \times 2^{6/5} \simeq 14.4, \tag{15}$$

which for GW150914 implies $M_{\rm max} \simeq 432\,\mathrm{M}_\odot$ (and $q \simeq 83$). This again forces the smaller mass to be at least $5\,\mathrm{M}_\odot$ – well above the neutron star mass limit (App. B).

The conclusion is the same as in the equal-mass or non-spinning case: both objects must be black holes.

## 4.4 Newtonian dynamics and compactness

We now examine the applicability of Newtonian dynamics. The dynamics will depart from the Newtonian approximation when the relative velocity $v$ approaches the speed of light or when the gravitational energy becomes large compared to the rest mass energy. For a binary system bound by gravity and with orbital velocity $v$, these two limits coincide and may be quantified by the post-Newtonian (PN) parameter [28] $x = (v/c)^2 = G\,M/\left(c^2\,r_{\rm sep}\right)$. Corrections to Newtonian dynamics may be expanded in powers of $x$, and are enumerated by their PN order. The 0PN approximation is precisely correct at $x = 0$, where dynamics are Newtonian and gravitational wave emission is described exactly by the quadrupole formula (Eq. 5).

The expression for the dimensionless PN parameter includes the Schwarzschild radius, so $x$ can be immediately recast in terms of the compactness ratio, $x \sim (2\mathscr{R})^{-1}$. As Newtonian dynamics holds when $x$ is small, the Newtonian approximation is valid down to compactness $\mathscr{R}$ of order of a few. Arguing by contradiction, if one assumes that the orbit is non-compact, then our analysis of the data using Newtonian mechanics is justified as an approximation of general relativity and leads to the conclusion that the orbit is compact.

If either of the bodies is rapidly spinning, their rotational velocity may also approach the speed of light, modifying the Newtonian dynamics, effectively adding spin-orbit and spin-spin interactions. However, these are also suppressed with a power of the PN parameter (1.5PN and 2PN, respectively [28–30]), and thus are significant only for compact orbits.

The same reasoning may also be applied to the use of the quadrupole formula [15] and/or to using the coordinate $R$ for the comparison of the Keplerian separation to the corresponding compact object radii (see Fig. 4 and its caption), as both of these are not entirely general and might be inaccurate. The separations are also subject to some arbitrariness due to gauge freedom. However here too, the errors in using these coordinates are non-negligible only in the orbits very close to a black hole, so again this argument does not refute our conclusions.

## 4.5 Is the chirp mass well measured? – constraints on the individual masses

As we are analyzing the final cycles before merger, having accepted that the bodies were compact, one might still ask whether Eq. 7 correctly describes the chirp mass in the non-Newtonian regime [31]. In fact for the last orbits, it does not: In Newtonian dynamics stable circular orbits may exist all the way down to merger, and energy lost to gravitational waves drives the inspiral between them. However in general relativity, close to the merger of compact objects (at least when one of the objects is much larger than the other) there are no such orbits past the innermost stable circular orbit (ISCO), whose typical location is given below. Allowed interior trajectories must be non-circular and "plunge" inwards (see pp. 911 of [18]). The changes in orbital separation and frequency in the final revolutions are thus not driven by the gravitational wave emission given by Eq. 7. This is why we used $f_{\rm GW}|_{\rm max}$ at the peak, rather than the final frequency $f_{\rm GW}|_{\rm fin}$.

We shall now constrain the individual masses based on $f_{\rm GW}|_{\rm fin}$, for which we do not need the Newtonian approximation at the late stage. No neutron stars have been observed above $3\,{\rm M}_\odot$; we shall rely on an even more conservative neutron star mass upper bound at $4.76\,{\rm M}_\odot$, a value chosen because given $\mathscr{M}$ from the early visible cycles, in order for the smaller mass $m_2$ to be below this threshold, $m_1$ must be at least $476\,{\rm M}_\odot$, which implies $q \geq 100$. Is such a high $q$ possible with the data that we have? Such a high mass ratio suggests a treatment of the system as an extremal mass ratio inspiral (EMRI), where the smaller mass approximately follows a geodesic orbit around the larger mass ($m_1 \sim M$). The frequencies of test-particle orbits (hence waveforms) around an object scale with the inverse of its mass, and also involve its dimensionless spin $\chi$. The orbital frequency $\omega_{\rm orb}$ as measured at infinity of a circular, equatorial orbit at radius $r$ (in Boyer-Lindquist coordinates) is given by [32]

$$\omega_{\rm orb} = \frac{\sqrt{GM}}{r^{3/2} + \chi\left(\sqrt{GM}/c\right)^3} = \frac{c^3}{GM}\left(\chi + \left(\frac{c^2 r}{GM}\right)^{3/2}\right)^{-1}. \tag{16}$$

For example, around a Schwarzschild black hole ($\chi = 0$) the quadrupole gravitational wave frequency at the innermost stable circular orbit (which is at $r = 6GM/c^2$) is hence equal to $f_{\rm GW} = 4.4({\rm M}_\odot/M)$ kHz, while for an extremal Kerr black hole ($\chi = 1$) the orbital frequency at ISCO ($r = GM/c^2$) is $\omega_{\rm orb} = c^3/2GM$, and the quadrupole gravitational frequency is $f_{\rm GW} = c^3/2\pi GM = 32({\rm M}_\odot/M)$ kHz. For a gravitational wave from the final plunge, the highest expected frequency is approximately the frequency from the light ring (LR), as nothing physical is expected to orbit faster than light,[8] and as waves originating within the light ring encounter an effective potential barrier at the light ring going out [33–37] . The light ring is at

$$r_{\rm LR} = \frac{2\,G\,M}{c^2}\left(1 + \cos\left(\frac{2}{3}\cos^{-1}(-\chi)\right)\right). \tag{17}$$

This radius is $3GM/c^2$ for a Schwarzschild black hole, while for a spinning Kerr black hole, as the spin $\chi$ increases the light ring radius decreases. For an extremal Kerr black hole it coincides with the innermost stable circular orbit at $GM/c^2$. The maximal gravitational wave frequency for a plunge into $m_1$ is then 67 Hz.

8 Hypothesized frequency up-conversions due to nonlinear GR effects have also been shown by NR to be absent [11–13].

Because we see gravitational wave emission from orbital motion at frequencies much higher than this maximal value, with or without spin, such a system is ruled out. Hence even the lighter of the masses must be at least $4.76 M_\odot > 3 M_\odot$, beyond the maximum observed mass of neutron stars.

### 4.6 Possible redshift of the masses – a constraint from the luminosity

Gravitational waves are stretched by the expansion of the Universe as they travel across it. This increases the wavelength and decreases the frequency of the waves observed on Earth compared to their values when emitted. The same effect accounts for the redshifting of photons from distant objects. The impact of this on the gravitational wave phasing corresponds to a scaling of the masses as measured on Earth; dimensional analysis of Eq. 7 shows that the source frame masses are smaller by $(1+z)$ relative to the detector frame, where $z$ is the redshift. Direct inspection of the detector data yields mass values from the red-shifted waves. How do these differ from their values at the source? In the next section, we estimate the distance to the source and hence the redshift, by relating the amplitude and luminosity of the gravitational wave from the merger to the observed strain and flux at the detector. The redshift is found to be $z \leq 0.1$, so the detector- and source-frame masses differ by less than of order 10%.

## 5 Luminosity and distance

Basic physics arguments also provide estimates of the peak gravitational wave luminosity of the system, its distance from us, and the total energy radiated in gravitational waves.

The gravitational wave amplitude $h$ falls off with increasing luminosity distance $d_L$ as $h \propto 1/d_L$. As shown in Fig. 1, the measured strain peaks at $h\big|_{\max} \sim 10^{-21}$. Had our detector been ten times closer to the source, the measured strain would have peaked at a value ten times larger. This could be continued, but the scaling relationship would break before $h$ reached unity, because near the Schwarzschild radius of the combined system $R \sim 200$ km the non-linear nature of gravity would become apparent. In this way we obtain a crude order-of-magnitude upper bound

$$d_L < 10^{21} \times 200 \text{ km} \sim 6 \text{ Gpc} \tag{18}$$

on the distance to the source.

We can obtain a more accurate distance estimate based on the luminosity, because the gravitational wave luminosity from an equal-mass binary inspiral has a peak value which is independent of the mass. This can be seen from naive dimensional analysis of the quadrupole formula, which gives a luminosity $L \sim \frac{G}{c^5} M^2 r^4 \omega^6$, with $\omega \sim c/r$ and $r \sim GM/c^2$, and $M\omega \sim c^3/G$ for the final tight orbit. Together this gives the Planck luminosity,[9]

$$L \sim L_{\text{Planck}} = c^5/G = 3.6 \times 10^{52} \text{ W}. \tag{19}$$

However, a closer look (Eq. A4) shows the prefactor should be $\frac{32}{5}\left(\frac{\mu}{M}\right)^2$, which gives $\frac{2}{5}$ for an equal-mass system, and is close to that for $q \sim 1$. Also, analysis of a small object falling into a Schwarzschild black hole suggests $M \sim \frac{1}{6} c^2 r_{\text{ISCO}}/G$ and $\omega r \sim 0.5c$. Taken together with the correct exponents, $L$ acquires a factor $0.4 \times 6^{-2} \times 0.5^6 \sim 0.2 \times 10^{-3}$. While the numerical value may change by a factor of a few with the specific spins, we can treat its order of magnitude as universal for similar-mass binaries.

Using Eq. 5 we relate the luminosity of gravitational waves to their strain $h$ at luminosity distance $d_L$,

$$L \sim \frac{c^3\, d_L^2}{4\,G} |\dot{h}|^2 \;\sim \frac{c^5}{4\,G} \left(\frac{\omega_{\text{GW}} d_L h}{c}\right)^2 . \tag{20}$$

Thus we have

$$\frac{L_{\text{peak}}}{L_{\text{Planck}}} \equiv \frac{L\big|_{\max}}{L_{\text{Planck}}} \sim 0.2 \times 10^{-3} \sim \left(\frac{\omega_{\text{GW}} d_L h\big|_{\max}}{c}\right)^2 , \tag{21}$$

and we estimate the distance from the change of the measured strain in time over the cycle at peak amplitude, as

$$d_L \sim 45\text{Gpc} \left(\frac{\text{Hz}}{f_{\text{GW}}\big|_{\max}}\right) \left(\frac{10^{-21}}{h\big|_{\max}}\right), \tag{22}$$

which for GW150914 gives $d_L \sim 300$ Mpc. This distance corresponds to a redshift of $z \leq 0.1$, and so does not substantially affect any of the conclusions. For a different distance-luminosity calculation based only on the strain data (reaching a similar estimate), see [42].

Using the orbital energy $E_{\text{orb}}$ (as defined in App. A) we may also estimate the total energy radiated as gravitational waves during the system's evolution from a very large initial separation (where $E^{\mathrm{i}}_{\mathrm{orb}} \to 0$) down to a separation $r$. For GW150914, using $m_1 \sim m_2 \sim 35\,\mathrm{M}_\odot$ and $r \sim R = 350\,\mathrm{km}$ (Eq. 9),

$$E_{\mathrm{GW}} = E^{\mathrm{i}}_{\mathrm{orb}} - E^{\mathrm{f}}_{\mathrm{orb}} = 0 - \left(-\frac{GM\mu}{2R}\right) \sim 3\,\mathrm{M}_\odot\, c^2. \qquad (23)$$

[9] The "Planck luminosity" $c^5/G$ has been proposed as the upper limit on the luminosity of any physical system [38–40]. Gibbons [41] has suggested that $c^5/4G$ be called the "Dyson luminosity" in honor of the physicist Freeman Dyson and because it is a *classical* quantity that does not contain the Planck constant $\hbar$.

This quantity should be considered an estimate for a lower bound on the total emitted energy (as some energy is emitted in the merger and ringdown); compare with the exact calculations in [1–3].

We note that the amount of energy emitted in this event is remarkable. During its ten-billion-year lifetime, our sun is expected to convert less than 1% of its mass into light and radiation. Not only did GW150914 release $\sim 300$ times as much energy in gravitational waves (almost entirely over the fraction of a second shown in Fig. 1), but for the cycle at peak luminosity, its power $L_{\mathrm{peak}}$ in the form of gravitational waves was about 22 orders of magnitude greater than the power output from our sun.

## 6 Conclusions

A lot of insight can be obtained by applying these basic physics arguments to the observed strain data of GW150914. These show the system that produced the gravitational wave was a pair of inspiraling black holes that approached very closely before merging. The system is seen to settle down, most likely to a single black hole. Simple arguments can also give us information about the system's distance and basic properties (for a related phenomenological approach see [43]).

With these basic arguments we have only drawn limited conclusions about the mass ratio $q$, because the frequency evolution described by Eq. 7 does not depend on $q$. The mass ratio $q$ does appear in the PN corrections [22, 44], thus its value can be further constrained from the data [2, 3].

These arguments will not work for every signal, for instance if the masses are too low to safely rule out a neutron star constituent as done in Sec. 4.5, but should be useful for systems similar to GW150914. There has already been another gravitational wave detection, GW151226 [6, 45], whose amplitude is smaller and therefore cannot be seen in the strain data without application of more advanced techniques.

Such techniques, combining analytic and numerical methods, can give us even more information, and we encourage the reader to explore how such analyses and models have been used for estimating the parameters of the system [2, 3], for testing and constraining the validity of general relativity in the highly relativistic, dynamic regime [4] and for astrophysical studies based on this event [5].

We hope that this paper will serve as an invitation to the field, at the beginning of the era of gravitational wave observations.

**Acknowledgements.** The authors gratefully acknowledge the support of the United States National Science Foundation (NSF) for the construction and operation of the LIGO Laboratory and Advanced LIGO as well as the Science and Technology Facilities Council (STFC) of the United Kingdom, the Max-Planck-Society (MPS), and the State of Niedersachsen/Germany for support of the construction of Advanced LIGO and construction and operation of the GEO600 detector. Additional support for Advanced LIGO was provided by the Australian Research Council. The authors gratefully acknowledge the Italian Istituto Nazionale di Fisica Nucleare (INFN), the French Centre National de la Recherche Scientifique (CNRS) and the Foundation for Fundamental Research on Matter supported by the Netherlands Organisation for Scientific Research, for the construction and operation of the Virgo detector and the creation and support of the EGO consortium. The authors also gratefully acknowledge research support from these agencies as well as by the Council of Scientific and Industrial Research of India, Department of Science and Technology, India, Science & Engineering Research Board (SERB), India, Ministry of Human Resource Development, India, the Spanish Ministerio de Economía y Competitividad, the Conselleria d'Economia i Competitivitat and Conselleria d'Educació, Cultura i Universitats of the Govern de les Illes Balears, the National Science Centre of Poland, the European Commission, the Royal Society, the Scottish Funding Council, the Scottish Universities Physics Alliance, the Hungarian Scientific Research Fund (OTKA), the Lyon Institute of Origins (LIO), the National Research Foundation of Korea, Industry Canada and the Province of Ontario through the Ministry of Economic Development and Innovation, the Natural Science and Engineering Research Council Canada, Canadian Institute for Advanced Research, the Brazilian Ministry of Science, Technology, and Innovation, Fundação de Amparo à Pesquisa do Estado de São Paulo (FAPESP), Russian Foundation for Basic Research, the Leverhulme Trust, the Research Corporation, Ministry of Science and Technology (MOST), Taiwan and the Kavli Foundation. The authors gratefully acknowledge the support of the NSF, STFC, MPS, INFN, CNRS and the State of Niedersachsen/Germany for provision of computational resources.

## Appendix A: Calculation of gravitational radiation from a binary system

Here we outline the calculation of the energy a binary system emits in gravitational waves and the emitted energy's effect on the system.

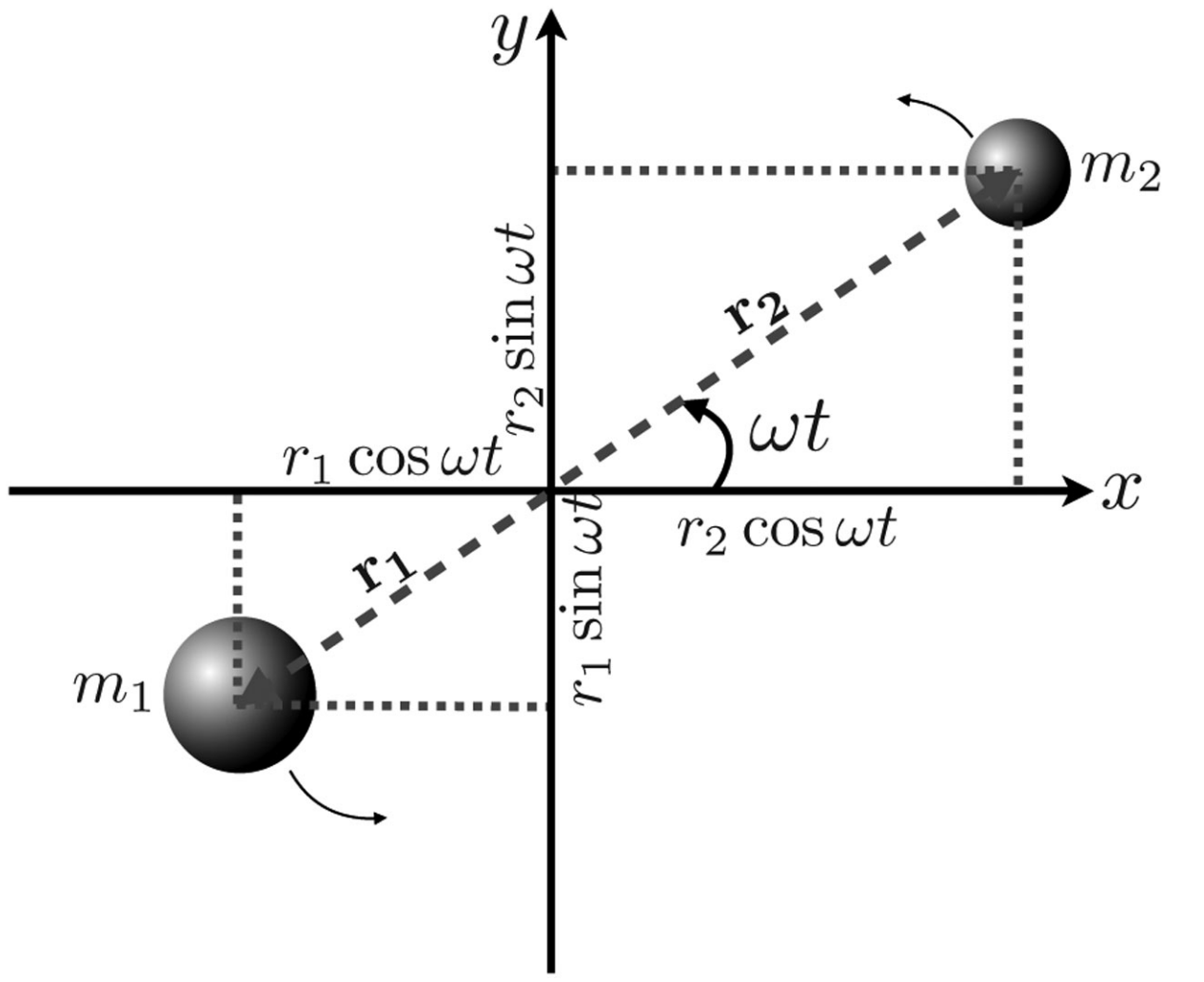

**Figure A1** A two-body system, $m_1$ and $m_2$ orbiting in the $xy$-plane around their C.O.M.

First we calculate the quadrupole moment $Q_{ij}$ of the system's mass distribution. We use a Cartesian coordinate system $\mathbf{x} = (x_1, x_2, x_3) = (x, y, z)$ whose origin is the center-of-mass, with $r$ the radial distance from the origin. $\delta_{ij} = \mathrm{diag}(1, 1, 1)$ is the Kronecker-delta and $\rho(\mathbf{x})$ denotes the mass density. Then

$$Q_{ij} = \int \mathrm{d}^3 x\, \rho(\mathbf{x}) \left( x_i x_j - \frac{1}{3} r^2 \delta_{ij} \right) \tag{A1}$$

$$= \sum_{A \in \{1,2\}} m_A \begin{pmatrix} \frac{2}{3}x_A^2 - \frac{1}{3}y_A^2 & x_A y_A & 0 \\ x_A y_A & \frac{2}{3}y_A^2 - \frac{1}{3}x_A^2 & 0 \\ 0 & 0 & -\frac{1}{3}r_A^2 \end{pmatrix}, \tag{A2}$$

where the second equality holds for a system of two bodies $A \in \{1, 2\}$ in the $xy$-plane. In the simple case of a circular orbit at separation $r = r_1 + r_2$ and frequency $f = \frac{\omega}{2\pi}$, a little trigonometry gives for each object (see Fig. A1)

$$Q_{ij}^A(t) = \frac{m_A r_A^2}{2} I_{ij}, \tag{A3}$$

where $I_{xx} = \cos(2\omega t) + \frac{1}{3}$, $I_{yy} = \frac{1}{3} - \cos(2\omega t)$, $I_{xy} = I_{yx} = \sin(2\omega t)$ and $I_{zz} = -\frac{2}{3}$. Combining we find $Q_{ij}(t) = \frac{1}{2}\mu r^2 I_{ij}$, and the gravitational wave luminosity from Eq. 5 is

$$\frac{\mathrm{d}}{\mathrm{d}t} E_{\mathrm{GW}} = \frac{32}{5} \frac{G}{c^5} \mu^2 r^4 \omega^6. \tag{A4}$$

This energy loss drains the orbital energy $E_{\mathrm{orb}} = -\frac{GM\mu}{2r}$, thus $\frac{\mathrm{d}}{\mathrm{d}t} E_{\mathrm{orb}} = \frac{GM\mu}{2r^2}\dot{r} = -\frac{\mathrm{d}}{\mathrm{d}t} E_{\mathrm{GW}}$. We assume that the energy radiated away over each orbit is small compared to $E_{\mathrm{orb}}$, in order to describe each orbit as approximately Keplerian.

Now using Kepler's third law $r^3 = GM/\omega^2$ and its derivative $\dot{r} = -\frac{2}{3} r\dot{\omega}/\omega$ we can substitute for all the $r$'s and obtain

$$\dot{\omega}^3 = \left(\frac{96}{5}\right)^3 \frac{\omega^{11}}{c^{15}} G^5 \mu^3 M^2 = \left(\frac{96}{5}\right)^3 \frac{\omega^{11}}{c^{15}} \left(G\mathscr{M}\right)^5, \tag{A5}$$

having defined the chirp mass $\mathscr{M} = \left(\mu^3 M^2\right)^{1/5}$.

We can see that Eq. A5 describes the evolution of the system as an inspiral: the orbital frequency goes up ("chirps"), while by Kepler's Law the orbital separation shrinks.

### A.1 Gravitational radiation from a different rotating system

A rising gravitational wave amplitude can accompany a rise in frequency in other rotating systems, evolving under different mechanisms. An increase in frequency means the system rotates faster and faster, so unless it gains angular momentum, the system's characteristic length $r(t)$ should be decreasing. For a system not driven by the loss of energy and angular momentum to gravitational waves, rapidly losing angular momentum is also difficult, thus the system should conserve its angular momentum $L = \alpha M r^2 \omega$, and so $\omega \propto L/r^2$.

The quadrupole formula (Eq. 4) then indicates the gravitational wave strain amplitude should follow the second time derivative of the quadrupole moment, $h \propto M r^2 w^2 \propto L\omega$.

Thus we see that for a system *not* driven by emission of gravitational waves, as the characteristic system size $r$ shrinks, both its gravitational wave frequency and amplitude grow, but remain proportional to each other. This is inconsistent with the data of GW150914 (Figs. 1, 2), which show the amplitude only grows by a factor of about 2 while the frequency $\omega(t)$ grows by at least a factor of 5.

## Appendix B: Possibilities for massive, compact objects

We are considering astrophysical objects with mass scale $m \sim 35\,\mathrm{M}_\odot$, which are constrained to fit into a radius $R$ such that the compactness ratio obeys $\mathscr{R} = \frac{c^2}{G}\frac{R}{m} \lesssim 3.4$. This produces a scale for their Newtonian density,

$$\rho \geq \frac{m}{(4\pi/3)\,R^3} = 3 \times 10^{15} \left(\frac{3.4}{\mathscr{R}}\right)^3 \left(\frac{35\,\mathrm{M}_\odot}{m}\right)^2 \frac{\mathrm{kg}}{\mathrm{m}^3}, \tag{B1}$$

where equality is attained for a uniform object. This is a factor of $10^6$ more dense than white dwarfs, so we can rule out objects supported by electron degeneracy pressure, as well as any main-sequence star, which would be less dense. While this density is a factor of $\sim 10^2$ *less* dense than neutron stars, these bodies exceed the maximum neutron star mass by an order of magnitude, as the neutron star limit is $\sim 3\mathrm{M}_\odot$ (3.2 $\mathrm{M}_\odot$ in [46, 47], 2.9 $\mathrm{M}_\odot$ in [48]). A more careful analysis of the frequency change, including tidal distortions, would have undoubtedly required the bodies to be even more compact in order to reach the final orbital frequency. This would push these massive bodies even closer to neutron-star density, thus constraining the equation of state into an even narrower corner. Thus, although theoretically a compactness ratio as low as $\mathscr{R} = 4/3$ is permitted for uniform objects [49], we can conclude that the data do show that if any of these objects were material bodies, they would need to occupy an extreme, narrow and heretofore unexplored and unobserved niche in the stellar continuum. The likeliest objects with such mass and compactness are black holes.

## Appendix C: Post inspiral phase: what we can conclude about the ringdown and the final object?

We have argued, using basic physics and scaling arguments, that the directly observable properties of the signal waveform for gravitational wave frequencies $f_{\mathrm{GW}} <$ 150 Hz shows that the source had been two black holes, which approached so closely that they subsequently merged. We now discuss the properties of the signal waveform at higher frequencies, and argue that this also lends support to this interpretation.

The data in Figures 1 and 2 show that after the peak gravitational wave amplitude is reached, the signal makes one to two additional cycles, continuing to rise in frequency until reaching about 250 Hz, while dropping sharply in amplitude. The frequency seems to level off just as the signal amplitude becomes hard to distinguish clearly.

Is this consistent with a merger remnant black hole? Immediately after being formed in a merger, a black hole horizon is very distorted. It proceeds to "lose its hair" and settle down to a final state of a Kerr black hole, uniquely defined [50] by its mass $M$ and spin parameter $\chi$. Late in this ringdown stage, the remaining perturbations should linearize, and the emitted gravitational wave should thus have characteristic quasi-normal-modes (QNMs). The set of QNMs is enumerated by various discrete indices, and their frequencies and damping times are determined by $M$ and $\chi$. Each such set would have a leading (least-damped) mode – and so finding a ringdown of several cycles with a fixed frequency would be strong evidence that a single final remnant was formed. We do clearly see the gravitational wave stabilizing in frequency (at around 250 Hz) about two cycles after the peak, and dying out in amplitude. Does the end of the observed waveform contain evidence of an exponentially-damped sinusoid of fixed frequency? Were such a mode found, analyzing its frequency and damping time, in conjunction with a model for black hole perturbations, could give an independent estimate of the mass and spin [51].

### C.1 Mode analysis

The ringing of a Kerr black hole can be thought of as related to a distortion of space-time traveling on a light ring orbit outside the black hole horizon (See [52] and references therein, and Eqs. (16, 17)); the expected frequency for a quadrupolar mode ($\ell = m = 2$) will thus be given as a dimensionless complex number

$$\frac{G}{c^3} M \omega_{\mathrm{GW}} = x + \mathrm{i} y. \tag{C1}$$

where the real part of $\omega_{\mathrm{GW}}$ is the angular frequency and the imaginary part is the (inverse) decay time. The ringdown amplitude and damping times are then found from

$$\mathrm{e}^{\mathrm{i}\omega_{\mathrm{GW}} t} = \mathrm{e}^{\mathrm{i}\frac{c^3 x}{GM} t}\,\mathrm{e}^{-\frac{c^3 y}{GM} t} = \mathrm{e}^{2\pi \mathrm{i} f_{\mathrm{GW}}|_{\mathrm{ringdown}} t}\,\mathrm{e}^{-t/\tau_{\mathrm{damp}}}\,, \tag{C2}$$

to be $f_{\mathrm{GW}}|_{\mathrm{ringdown}} = c^3 x / (2\pi GM)$ and $\tau_{\mathrm{damp}} = GM/c^3 y$.

The exact values of $x$ and $y$ can be found as when analyzing the normal modes of a resonant cavity: one uses separation of variables to solve the field equations, and then enforces the boundary conditions to obtain a discrete set of complex eigenfrequencies [52]. However, limiting values on $x$, $x \in (\sim 0.3, 1]$, are derived immediately from Eqs. (16, 17), with a factor of 2 between orbital and gravitational wave frequencies. The final gravitational wave frequency is thus determined by the mass (up to the order-of-unity factor $x$, which embodies the spin). We have in fact already used this to show how our high attained frequency constrains the total mass and the compactness of the objects (objects of larger radius would have distortion bulges orbiting much farther than the light ring, mandating much lower frequencies). For the parameter $y$ determining the damping time, numerical tabulations of the QNMs [52] show that

$$f_{\mathrm{GW}}|_{\mathrm{ringdown}}\, \tau_{\mathrm{damp}} = \frac{x}{2\pi y} \sim 1 \tag{C3}$$

for a broad range of mode numbers and spins, as long as the spin is not close to extremal. This shows that the ringdown is expected to have a damping time roughly equal to the period of oscillation. This is exactly what is seen in the waveform, and is the reason the signal amplitude drops so low by the time the remnant rings at the final frequency.

While it is beyond the scope of this paper to calculate the exact QNMs for black holes of different spins, or to find the final spin of a general black hole merger, it is worth mentioning that for a wide range of spins for similar-mass binaries, the final spin is expected to be about $\chi \sim 0.7$, for which Eq. (16, 17) estimate that $\mathrm{Re}[\frac{G}{c^3} M\omega_{\mathrm{GW}}] \sim 0.55$.

The exact value can be found using Table II in [52], where the leading harmonic ($\ell = 2$, $m = 2$, $n = 0$) for a black hole with a spin $\chi = 0.7$ has $\frac{G}{c^3} M\omega_{\mathrm{GW}} = 0.5326 + 0.0808i$, giving a ringdown frequency

$$f_{\mathrm{GW}}|_{\mathrm{ringdown}} \approx 260\mathrm{Hz} \left( \frac{65\,\mathrm{M}_\odot}{M} \right) , \tag{C4}$$

and a damping time

$$\tau_{\mathrm{damp}} = 4\ \mathrm{ms} \left( \frac{M}{65\,\mathrm{M}_\odot} \right) \sim \frac{1}{f_{\mathrm{GW}}|_{\mathrm{ringdown}}}. \tag{C5}$$

In other words, the signal in the data is fully consistent [42] with the final object being a Kerr black hole with a dimensionless spin parameter $\chi \sim 0.7$ and a mass $M \sim 65\,\mathrm{M}_\odot$ Such a final mass is consistent with the merger of two black holes of $\sim 35\,\mathrm{M}_\odot$ each, after accounting for the energy emitted as gravitational waves (Eq. 23). This interpretation of the late part of the signal is also consistent with numerical simulations [53]. Full numerical simulations from the peak and onward, where the signal amplitude is considerably higher, also show consistency with the formation of a Kerr black hole remnant [2, 4].

**Key words.** GW150914, gravitational waves, black holes.

## Authors

*B. P. Abbott[1], R. Abbott[1], T. D. Abbott[2], M. R. Abernathy[3], F. Acernese[4,5], K. Ackley[6], C. Adams[7], T. Adams[8], P. Addesso[9], R. X. Adhikari[1], V. B. Adya[10], C. Affeldt[10], M. Agathos[11], K. Agatsuma[11], N. Aggarwal[12], O. D. Aguiar[13], L. Aiello[14,15], A. Ain[16], P. Ajith[17], B. Allen[10,18,19], A. Allocca[20,21], P. A. Altin[22], S. B. Anderson[1], W. G. Anderson[18], K. Arai[1], M. C. Araya[1], C. C. Arceneaux[23], J. S. Areeda[24], N. Arnaud[25], K. G. Arun[26], S. Ascenzi[27,15], G. Ashton[28], M. Ast[29], S. M. Aston[7], P. Astone[30], P. Aufmuth[19], C. Aulbert[10], S. Babak[31], P. Bacon[32], M. K. M. Bader[11], F. Baldaccini[33,34], G. Ballardin[35], S. W. Ballmer[36], J. C. Barayoga[1], S. E. Barclay[37], B. C. Barish[1], D. Barker[38], F. Barone[4,5], B. Barr[37], L. Barsotti[12], M. Barsuglia[32], D. Barta[39], J. Bartlett[38], I. Bartos[40], R. Bassiri[41], A. Basti[20,21], J. C. Batch[38], C. Baune[10], V. Bavigadda[35], M. Bazzan[42,43], M. Bejger[44], A. S. Bell[37], G. Bergmann[10], C. P. L. Berry[45], D. Bersanetti[46,47], A. Bertolini[11], J. Betzwieser[7], S. Bhagwat[36], R. Bhandare[48], I. A. Bilenko[49], G. Billingsley[1], J. Birch[7], R. Birney[50], O. Birnholtz[10], S. Biscans[12], A. Bisht[10,19], M. Bitossi[35], C. Biwer[36], M. A. Bizouard[25], J. K. Blackburn[1], C. D. Blair[51], D. G. Blair[51], R. M. Blair[38], S. Bloemen[52], O. Bock[10], M. Boer[53], G. Bogaert[53], C. Bogan[10], A. Bohe[31], C. Bond[45], F. Bondu[54], R. Bonnand[8], B. A. Boom[11], R. Bork[1], V. Boschi[20,21], S. Bose[55,16], Y. Bouffanais[32], A. Bozzi[35], C. Bradaschia[21], V. B. Braginsky[49,∗∗∗], M. Branchesi[56,57], J. E. Brau[58], T. Briant[59], A. Brillet[53], M. Brinkmann[10], V. Brisson[25], P. Brockill[18], J. E. Broida[60], A. F. Brooks[1], D. A. Brown[36], D. D. Brown[45], N. M. Brown[12], S. Brunett[1], C. C. Buchanan[2], A. Buikema[12], T. Bulik[61], H. J. Bulten[62,11], A. Buonanno[31,63], D. Buskulic[8], C. Buy[32], R. L. Byer[41], M. Cabero[10], L. Cadonati[64], G. Cagnoli[65,66], C. Cahillane[1], J. Calderón Bustillo[64], T. Callister[1], E. Calloni[67,5], J. B. Camp[68], K. C. Cannon[69], J. Cao[70], C. D. Capano[10], E. Capocasa[32], F. Carbognani[35], S. Caride[71], J. Casanueva Diaz[25], C. Casentini[27,15], S. Caudill[18], M. Cavaglià[23], F. Cavalier[25], R. Cavalieri[35], G. Cella[21], C. B. Cepeda[1], L. Cerboni Baiardi[56,57], G. Cerretani[20,21], E. Cesarini[27,15], S. J. Chamberlin[72], M. Chan[37], S. Chao[73], P. Charlton[74], E. Chassande-Mottin[32], H. Y. Chen[75], Y. Chen[76], C. Cheng[73], A. Chincarini[47], A. Chiummo[35], H. S. Cho[77], M. Cho[63], J. H. Chow[22], N. Christensen[60], Q. Chu[51], S. Chua[59], S. Chung[51], G. Ciani[6], F. Clara[38], J. A. Clark[64], F. Cleva[53], E. Coccia[27,14], P.-F. Cohadon[59], A. Colla[78,30], C. G. Collette[79], L. Cominsky[80], M. Constancio Jr.[13], A. Conte[78,30], L. Conti[43], D. Cook[38], T. R. Corbitt[2], A. Corsi[71], S. Cortese[35], C. A. Costa[13], M. W. Coughlin[60], S. B. Coughlin[81], J.-P. Coulon[53], S. T. Countryman[40], P. Couvares[1], E. E. Cowan[64], D. M. Coward[51], M. J. Cowart[7], D. C. Coyne[1], R. Coyne[71], K. Craig[37], J. D. E. Creighton[18], J. Cripe[2], S. G. Crowder[82], A. Cumming[37], L. Cunningham[37], E. Cuoco[35], T. Dal Canton[10], S. L. Danilishin[37], S. D'Antonio[15], K. Danzmann[19,10], N. S. Darman[83], A. Dasgupta[84], C. F. Da Silva Costa[6], V. Dattilo[35], I. Dave[48], M. Davier[25], G. S. Davies[37], E. J. Daw[85], R. Day[35], S. De[36], D. DeBra[41], G. Debreczeni[39], J. Degallaix[65], M. De Laurentis[67,5], S. Deléglise[59], W. Del Pozzo[45], T. Denker[10], T. Dent[10], V. Dergachev[1], R. De Rosa[67,5], R. T. DeRosa[7], R. DeSalvo[9], R. C. Devine[86], S. Dhurandhar[16], M. C. Díaz[87], L. Di Fiore[5], M. Di Giovanni[88,89], T. Di Girolamo[67,5], A. Di Lieto[20,21], S. Di Pace[78,30], I. Di Palma[31,78,30], A. Di Virgilio[21], V. Dolique[65], F. Donovan[12], K. L. Dooley[23], S. Doravari[10], R. Douglas[37], T. P. Downes[18], M. Drago[10], R. W. P. Drever[1], J. C. Driggers[38], M. Ducrot[8], S. E. Dwyer[38], T. B. Edo[85], M. C. Edwards[60], A. Effler[7], H.-B. Eggenstein[10], P. Ehrens[1], J. Eichholz[6,1], S. S. Eikenberry[6], W. Engels[76], R. C. Essick[12], T. Etzel[1], M. Evans[12], T. M. Evans[7], R. Everett[72], M. Factourovich[40], V. Fafone[27,15], H. Fair[36], S. Fairhurst[90], X. Fan[70], Q. Fang[51], S. Farinon[47], B. Farr[75], W. M. Farr[45], M. Favata[91], M. Fays[90], H. Fehrmann[10], M. M. Fejer[41], E. Fenyvesi[92], I. Ferrante[20,21], E. C. Ferreira[13], F. Ferrini[35], F. Fidecaro[20,21], I. Fiori[35], D. Fiorucci[32], R. P. Fisher[36], R. Flaminio[65,93], M. Fletcher[37], J.-D. Fournier[53], S. Frasca[78,30], F. Frasconi[21], Z. Frei[92], A. Freise[45], R. Frey[58], V. Frey[25], P. Fritschel[12], V. V. Frolov[7], P. Fulda[6], M. Fyffe[7], H. A. G. Gabbard[23], J. R. Gair[94], L. Gammaitoni[33], S. G. Gaonkar[16], F. Garufi[67,5], G. Gaur[95,84], N. Gehrels[68], G. Gemme[47], P. Geng[87], E. Genin[35], A. Gennai[21], J. George[48], L. Gergely[96], V. Germain[8], Abhirup Ghosh[17], Archisman Ghosh[17], S. Ghosh[52,11], J. A. Giaime[2,7], K. D. Giardina[7], A. Giazotto[21], K. Gill[97], A. Glaefke[37], E. Goetz[38], R. Goetz[6], L. Gondan[92], G. González[2], J. M. Gonzalez Castro[20,21], A. Gopakumar[98], N. A. Gordon[37], M. L. Gorodetsky[49], S. E. Gossan[1], M. Gosselin[35], R. Gouaty[8], A. Grado[99,5], C. Graef[37], P. B. Graff[63], M. Granata[65], A. Grant[37], S. Gras[12], C. Gray[38], G. Greco[56,57], A. C. Green[45], P. Groot[52], H. Grote[10], S. Grunewald[31], G. M. Guidi[56,57], X. Guo[70], A. Gupta[16], M. K. Gupta[84], K. E. Gushwa[1], E. K. Gustafson[1], R. Gustafson[100], J. J. Hacker[24], B. R. Hall[55], E. D. Hall[1], G. Hammond[37], M. Haney[98], M. M. Hanke[10], J. Hanks[38], C. Hanna[72], M. D. Hannam[90], J. Hanson[7], T. Hardwick[2], J. Harms[56,57], G. M. Harry[3], I. W. Harry[31], M. J. Hart[37], M. T. Hartman[6], C.-J. Haster[45], K. Haughian[37], A. Heidmann[59], M. C. Heintze[7], H. Heitmann[53], P. Hello[25], G. Hemming[35],*

M. Hendry[37], I. S. Heng[37], J. Hennig[37], J. Henry[101], A. W. Heptonstall[1], M. Heurs[10,19], S. Hild[37], D. Hoak[35], D. Hofman[65], K. Holt[7], D. E. Holz[75], P. Hopkins[90], J. Hough[37], E. A. Houston[37], E. J. Howell[51], Y. M. Hu[10], S. Huang[73], E. A. Huerta[102], D. Huet[25], B. Hughey[97], S. Husa[103], S. H. Huttner[37], T. Huynh-Dinh[7], N. Indik[10], D. R. Ingram[38], R. Inta[71], H. N. Isa[37], J.-M. Isac[59], M. Isi[1], T. Isogai[12], B. R. Iyer[17], K. Izumi[38], T. Jacqmin[59], H. Jang[77], K. Jani[64], P. Jaranowski[104], S. Jawahar[105], L. Jian[51], F. Jiménez-Forteza[103], W. W. Johnson[2], D. I. Jones[28], R. Jones[37], R. J. G. Jonker[11], L. Ju[51], Haris K[106], C. V. Kalaghatgi[90], V. Kalogera[81], S. Kandhasamy[23], G. Kang[77], J. B. Kanner[1], S. J. Kapadia[10], S. Karki[58], K. S. Karvinen[10], M. Kasprzack[35,2], E. Katsavounidis[12], W. Katzman[7], S. Kaufer[19], T. Kaur[51], K. Kawabe[38], F. Kéfélian[53], M. S. Kehl[107], D. Keitel[103], D. B. Kelley[36], W. Kells[1], R. Kennedy[85], J. S. Key[87], F. Y. Khalili[49], I. Khan[14], S. Khan[90], Z. Khan[84], E. A. Khazanov[108], N. Kijbunchoo[38], Chi-Woong Kim[77], Chunglee Kim[77], J. Kim[109], K. Kim[110], N. Kim[41], W. Kim[111], Y.-M. Kim[109], S. J. Kimbrell[64], E. J. King[111], P. J. King[38], J. S. Kissel[38], B. Klein[81], L. Kleybolte[29], S. Klimenko[6], S. M. Koehlenbeck[10], S. Koley[11], V. Kondrashov[1], A. Kontos[12], M. Korobko[29], W. Z. Korth[1], I. Kowalska[61], D. B. Kozak[1], V. Kringel[10], B. Krishnan[10], A. Królak[112,113], C. Krueger[19], G. Kuehn[10], P. Kumar[107], R. Kumar[84], L. Kuo[73], A. Kutynia[112], B. D. Lackey[36], M. Landry[38], J. Lange[101], B. Lantz[41], P. D. Lasky[114], M. Laxen[7], A. Lazzarini[1], C. Lazzaro[43], P. Leaci[78,30], S. Leavey[37], E. O. Lebigot[32,70], C. H. Lee[109], H. K. Lee[110], H. M. Lee[115], K. Lee[37], A. Lenon[36], M. Leonardi[88,89], J. R. Leong[10], N. Leroy[25], N. Letendre[8], Y. Levin[114], J. B. Lewis[1], T. G. F. Li[116], A. Libson[12], T. B. Littenberg[117], N. A. Lockerbie[105], A. L. Lombardi[118], L. T. London[90], J. E. Lord[36], M. Lorenzini[14,15], V. Loriette[119], M. Lormand[7], G. Losurdo[57], J. D. Lough[10,19], H. Lück[19,10], A. P. Lundgren[10], R. Lynch[12], Y. Ma[51], B. Machenschalk[10], M. MacInnis[12], D. M. Macleod[2], F. Magaña-Sandoval[36], L. Magaña Zertuche[36], R. M. Magee[55], E. Majorana[30], I. Maksimovic[119], V. Malvezzi[27,15], N. Man[53], V. Mandic[82], V. Mangano[37], G. L. Mansell[22], M. Manske[18], M. Mantovani[35], F. Marchesoni[120,34], F. Marion[8], S. Márka[40], Z. Márka[40], A. S. Markosyan[41], E. Maros[1], F. Martelli[56,57], L. Martellini[53], I. W. Martin[37], D. V. Martynov[12], J. N. Marx[1], K. Mason[12], A. Masserot[8], T. J. Massinger[36], M. Masso-Reid[37], S. Mastrogiovanni[78,30], F. Matichard[12], L. Matone[40], N. Mavalvala[12], N. Mazumder[55], R. McCarthy[38], D. E. McClelland[22], S. McCormick[7], S. C. McGuire[121], G. McIntyre[1], J. McIver[1], D. J. McManus[22], T. McRae[22], D. Meacher[72], G. D. Meadors[31,10], J. Meidam[11], A. Melatos[83], G. Mendell[38], R. A. Mercer[18], E. L. Merilh[38], M. Merzougui[53], S. Meshkov[1], C. Messenger[37], C. Messick[72], R. Metzdorff[59], P. M. Meyers[82], F. Mezzani[30,78], H. Miao[45], C. Michel[65], H. Middleton[45], E. E. Mikhailov[122], L. Milano[67,5], A. L. Miller[6,78,30], A. Miller[81], B. B. Miller[81], J. Miller[12], M. Millhouse[123], Y. Minenkov[15], J. Ming[31], S. Mirshekari[124], C. Mishra[17], S. Mitra[16], V. P. Mitrofanov[49], G. Mitselmakher[6], R. Mittleman[12], A. Moggi[21], M. Mohan[35], S. R. P. Mohapatra[12], M. Montani[56,57], B. C. Moore[91], C. J. Moore[125], D. Moraru[38], G. Moreno[38], S. R. Morriss[87], K. Mossavi[10], B. Mours[8], C. M. Mow-Lowry[45], G. Mueller[6], A. W. Muir[90], Arunava Mukherjee[17], D. Mukherjee[18], S. Mukherjee[87], N. Mukund[16], A. Mullavey[7], J. Munch[111], D. J. Murphy[40], P. G. Murray[37], A. Mytidis[6], I. Nardecchia[27,15], L. Naticchioni[78,30], R. K. Nayak[126], K. Nedkova[118], G. Nelemans[52,11], T. J. N. Nelson[7], M. Neri[46,47], A. Neunzert[100], G. Newton[37], T. T. Nguyen[22], A. B. Nielsen[10], S. Nissanke[52,11], A. Nitz[10], F. Nocera[35], D. Nolting[7], M. E. N. Normandin[87], L. K. Nuttall[36], J. Oberling[38], E. Ochsner[18], J. O'Dell[127], E. Oelker[12], G. H. Ogin[128], J. J. Oh[129], S. H. Oh[129], F. Ohme[90], M. Oliver[103], P. Oppermann[10], Richard J. Oram[7], B. O'Reilly[7], R. O'Shaughnessy[101], D. J. Ottaway[111], H. Overmier[7], B. J. Owen[71], A. Pai[106], S. A. Pai[48], J. R. Palamos[58], O. Palashov[108], C. Palomba[30], A. Pal-Singh[29], H. Pan[73], C. Pankow[81], F. Pannarale[90], B. C. Pant[48], F. Paoletti[35,21], A. Paoli[35], M. A. Papa[31,18,10], H. R. Paris[41], W. Parker[7], D. Pascucci[37], A. Pasqualetti[35], R. Passaquieti[20,21], D. Passuello[21], B. Patricelli[20,21], Z. Patrick[41], B. L. Pearlstone[37], M. Pedraza[1], R. Pedurand[65,130], L. Pekowsky[36], A. Pele[7], S. Penn[131], A. Perreca[1], L. M. Perri[81], M. Phelps[37], O. J. Piccinni[78,30], M. Pichot[53], F. Piergiovanni[56,57], V. Pierro[9], G. Pillant[35], L. Pinard[65], I. M. Pinto[9], M. Pitkin[37], M. Poe[18], R. Poggiani[20,21], P. Popolizio[35], A. Post[10], J. Powell[37], J. Prasad[16], J. Pratt[97], V. Predoi[90], T. Prestegard[82], L. R. Price[1], M. Prijatelj[10,35], M. Principe[9], S. Privitera[31], R. Prix[10], G. A. Prodi[88,89], L. Prokhorov[49], O. Puncken[10], M. Punturo[34], P. Puppo[30], M. Pürrer[31], H. Qi[18], J. Qin[51], S. Qiu[114], V. Quetschke[87], E. A. Quintero[1], R. Quitzow-James[58], F. J. Raab[38], D. S. Rabeling[22], H. Radkins[38], P. Raffai[92], S. Raja[48], C. Rajan[48], M. Rakhmanov[87], P. Rapagnani[78,30], V. Raymond[31], M. Razzano[20,21], V. Re[27], J. Read[24], C. M. Reed[38], T. Regimbau[53], L. Rei[47], S. Reid[50], H. Rew[122], S. D. Reyes[36], F. Ricci[78,30], K. Riles[100], M. Rizzo[101], N. A. Robertson[1,37], R. Robie[37], F. Robinet[25], A. Rocchi[15], L. Rolland[8], J. G. Rollins[1], V. J. Roma[58], J. D. Romano[87], R. Romano[4,5], G. Romanov[122], J. H. Romie[7], D. Rosińska[132,44], S. Rowan[37], A. Rüdiger[10], P. Ruggi[35], K. Ryan[38], S. Sachdev[1], T. Sadecki[38], L. Sadeghian[18], M. Sakellariadou[133], L. Salconi[35], M. Saleem[106], F. Salemi[10], A. Samajdar[126], L. Sammut[114], E. J. Sanchez[1], V. Sandberg[38], B. Sandeen[81], J. R. Sanders[36], B. Sassolas[65], P. R. Saulson[36], O. E. S. Sauter[100], R. L. Savage[38], A. Sawadsky[19], P. Schale[58], R. Schilling[10,§], J. Schmidt[10], P. Schmidt[1,76], R. Schnabel[29], R. M. S. Schofield[58], A. Schönbeck[29], E. Schreiber[10], D. Schuette[10,19], B. F. Schutz[90,31], J. Scott[37], S. M. Scott[22], D. Sellers[7], A. S. Sengupta[95], D. Sentenac[35], V. Sequino[27,15], A. Sergeev[108], Y. Setyawati[52,11], D. A. Shaddock[22], T. Shaffer[38], M. S. Shahriar[81], M. Shaltev[10], B. Shapiro[41], P. Shawhan[63], A. Sheperd[18], D. H. Shoemaker[12], D. M. Shoemaker[64], K. Siellez[64], X. Siemens[18], M. Sieniawska[44], D. Sigg[38], A. D. Silva[13], A. Singer[1], L. P. Singer[68], A. Singh[31,10,19], R. Singh[2], A. Singhal[14], A. M. Sintes[103], B. J. J. Slagmolen[22], J. R. Smith[24], N. D. Smith[1], R. J. E. Smith[1], E. J. Son[129], B. Sorazu[37], F. Sorrentino[47], T. Souradeep[16], A. K. Srivastava[84], A. Staley[40], M. Steinke[10], J. Steinlechner[37], S. Steinlechner[37], D. Steinmeyer[10,19], B. C. Stephens[18], R. Stone[87], K. A. Strain[37], N. Straniero[65], G. Stratta[56,57], N. A. Strauss[60], S. Strigin[49], R. Sturani[124], A. L. Stuver[7], T. Z. Summerscales[134], L. Sun[83], S. Sunil[84], P. J. Sutton[90], B. L. Swinkels[35], M. J. Szczepańczyk[97], M. Tacca[32], D. Talukder[58], D. B. Tanner[6], M. Tápai[96], S. P. Tarabrin[10], A. Taracchini[31], R. Taylor[1], T. Theeg[10], M. P. Thirugnanasambandam[1], E. G. Thomas[45], M. Thomas[7], P. Thomas[38], K. A. Thorne[7], K. S. Thorne[76], E. Thrane[114], S. Tiwari[14,89], V. Tiwari[90], K. V. Tokmakov[105], K. Toland[37], C. Tomlinson[85], M. Tonelli[20,21], Z. Tornasi[37], C. V. Torres[87,‖], C. I. Torrie[1], D. Töyrä[45], F. Travasso[33,34], G. Traylor[7], D. Trifirò[23], M. C. Tringali[88,89], L. Trozzo[135,21], M. Tse[12], M. Turconi[53], D. Tuyenbayev[87], D. Ugolini[136], C. S. Unnikrishnan[98], A. L. Urban[18], S. A. Usman[36], H. Vahlbruch[19], G. Vajente[1], G. Valdes[87], N. van Bakel[11], M. van Beuzekom[11], J. F. J. van den Brand[62,11], C. Van Den Broeck[11], D. C. Vander-Hyde[36], L. van der Schaaf[11], J. V. van Heijningen[11], A. A. van Veggel[37], M. Vardaro[42,43], S. Vass[1], M. Vasúth[39], R. Vaulin[12], A. Vecchio[45], G. Vedovato[43], J. Veitch[45], P. J. Veitch[111], K. Venkateswara[137], D. Verkindt[8], F. Vetrano[56,57], A. Viceré[56,57], S. Vinciguerra[45], D. J. Vine[50], J.-Y. Vinet[53], S. Vitale[12], T. Vo[36], H. Vocca[33,34], C. Vorvick[38], D. V. Voss[6], W. D. Vousden[45], S. P. Vyatchanin[49], A. R. Wade[22], L. E. Wade[138], M. Wade[138], M. Walker[2], L. Wallace[1], S. Walsh[31,10], G. Wang[14,57], H. Wang[45], M. Wang[45], X. Wang[70], Y. Wang[51], R. L. Ward[22], J. Warner[38], M. Was[8], B. Weaver[38], L.-W. Wei[53], M. Weinert[10], A. J. Weinstein[1], R. Weiss[12], L. Wen[51], P. Weßels[10], T. Westphal[10], K. Wette[10], J. T. Whelan[101], B. F. Whiting[6], R. D. Williams[1], A. R. Williamson[90], J. L. Willis[139], B. Willke[19,10], M. H. Wimmer[10,19], W. Winkler[10], C. C. Wipf[1], A. G. Wiseman[18], H. Wittel[10,19], G. Woan[37], J. Woehler[10], J. Worden[38], J. L. Wright[37], D. S. Wu[10], G. Wu[7], J. Yablon[81],

Original Paper

*W. Yam*[12], *H. Yamamoto*[1], *C. C. Yancey*[63], *H. Yu*[12], *M. Yvert*[8], *A. Zadrożny*[112], *L. Zangrando*[43], *M. Zanolin*[97], *J.-P. Zendri*[43], *M. Zevin*[81], *L. Zhang*[1], *M. Zhang*[122], *Y. Zhang*[101], *C. Zhao*[51], *M. Zhou*[81], *Z. Zhou*[81], *X. J. Zhu*[51], *M. E. Zucker*[1,12], *S. E. Zuraw*[118], *and J. Zweizig*[1]

[***]Deceased, March 2016.
[§]Deceased, May 2015.
[‖]Deceased, March 2015.

[1]LIGO, California Institute of Technology, Pasadena, CA 91125, USA
[2]Louisiana State University, Baton Rouge, LA 70803, USA
[3]American University, Washington, D.C., 20016, USA
[4]Università di Salerno, Fisciano, I-84084 Salerno, Italy
[5]INFN, Sezione di Napoli, Complesso Universitario di Monte S.Angelo, I-80126 Napoli, Italy
[6]University of Florida, Gainesville FL, 32611, USA
[7]LIGO Livingston Observatory, Livingston, LA 70754, USA
[8]Laboratoire d'Annecy-le-Vieux de Physique des Particules (LAPP), Université Savoie Mont Blanc, CNRS/IN2P3, F-74941 Annecy-le-Vieux, France
[9]University of Sannio at Benevento, I-82100 Benevento, Italy and INFN, Sezione di Napoli, I-80100, Napoli Italy
[10]Albert-Einstein-Institut, Max-Planck-Institut für Gravitationsphysik, D-30167, Hannover Germany
[11]Nikhef, Science Park, 1098 XG, Amsterdam, The Netherlands
[12]LIGO, Massachusetts Institute of Technology, Cambridge, MA 02139, USA
[13]Instituto Nacional de Pesquisas Espaciais, 12227-010, São José dos Campos, São Paulo, Brazil
[14]INFN, Gran Sasso Science Institute, I-67100 L'Aquila, Italy
[15]INFN, Sezione di Roma Tor Vergata, I-00133 Roma Italy
[16]Inter-University Centre for Astronomy and Astrophysics, Pune 411007, India
[17]International Centre for Theoretical Sciences, Tata Institute of Fundamental Research, Bangalore 560012, India
[18]University of Wisconsin-Milwaukee, Milwaukee, WI 53201 USA
[19]Leibniz Universität Hannover, D-30167 Hannover, Germany
[20]Università di Pisa, I-56127 Pisa, Italy
[21]INFN, Sezione di Pisa, I-56127 Pisa, Italy
[22]Australian National University, Canberra, Australian Capital Territory 0200, Australia
[23]The University of Mississippi, University, MS 38677, USA
[24]California State University Fullerton, Fullerton, CA 92831, USA
[25]LAL, Univ. Paris-Sud, CNRS/IN2P3, Université Paris-Saclay, Orsay, France
[26]Chennai Mathematical Institute, Chennai 603103, India
[27]Università di Roma Tor Vergata, I-00133 Roma, Italy
[28]University of Southampton, Southampton SO17 1BJ, United Kingdom
[29]Universität Hamburg, D-22761 Hamburg, Germany
[30]INFN, Sezione di Roma, I-00185 Roma, Italy
[31]Albert-Einstein-Institut, Max-Planck-Institut für Gravitationsphysik, D-14476 Potsdam-Golm, Germany
[32]APC, AstroParticule et Cosmologie, Université Paris Diderot, CNRS/IN2P3, CEA/Irfu, Observatoire de Paris, Sorbonne Paris Cité, F-75205 Paris Cedex 13, France
[33]Università di Perugia, I-06123 Perugia, Italy
[34]INFN, Sezione di Perugia, I-06123 Perugia, Italy
[35]European Gravitational Observatory (EGO), I-56021 Cascina, Pisa, Italy
[36]Syracuse University, Syracuse, NY 13244, USA
[37]SUPA, University of Glasgow, Glasgow G12 8QQ, United Kingdom
[38]LIGO Hanford Observatory, Richland, WA 99352, USA
[39]Wigner RCP, RMKI, H-1121, Budapest, Konkoly Thege Miklós út 29-33, Hungary
[40]Columbia University, New York, NY 10027, USA
[41]Stanford University, Stanford, CA 94305, USA
[42]Università di Padova, Dipartimento di Fisica e Astronomia, I-35131 Padova, Italy
[43]INFN, Sezione di Padova, I-35131 Padova, Italy
[44]CAMK-PAN, 00-716 Warsaw, Poland
[45]University of Birmingham, Birmingham B15 2TT, United Kingdom
[46]Università degli Studi di Genova, I-16146 Genova, Italy
[47]INFN, Sezione di Genova, I-16146 Genova, Italy
[48]RRCAT, Indore, MP 452013, India
[49]Faculty of Physics, Lomonosov Moscow State University, Moscow 119991, Russia
[50]SUPA, University of the West of Scotland, Paisley PA1 2BE, United Kingdom
[51]University of Western Australia, Crawley, Western Australia 6009, Australia
[52]Department of Astrophysics/IMAPP, Radboud University Nijmegen, P.O. Box 9010, 6500 GL, Nijmegen, The Netherlands
[53]Artemis, Université Côte d'Azur, CNRS, Observatoire Côte d'Azur, CS 34229, Nice cedex 4, France
[54]Institut de Physique de Rennes, CNRS, Université de Rennes 1, F-35042 Rennes, France
[55]Washington State University, Pullman, WA 99164, USA
[56]Università degli Studi di Urbino "Carlo Bo", I-61029 Urbino, Italy
[57]INFN, Sezione di Firenze, I-50019 Sesto Fiorentino, Firenze, Italy
[58]University of Oregon, Eugene, OR 97403, USA
[59]Laboratoire Kastler Brossel, UPMC-Sorbonne Universités, CNRS, ENS-PSL Research University, Collège de France, F-75005 Paris, France
[60]Carleton College, Northfield, MN 55057, USA
[61]Astronomical Observatory Warsaw University, 00-478 Warsaw, Poland
[62]VU University Amsterdam, 1081 HV, Amsterdam, The Netherlands
[63]University of Maryland, College Park, MD 20742, USA
[64]Center for Relativistic Astrophysics and School of Physics, Georgia Institute of Technology, Atlanta, GA 30332, USA
[65]Laboratoire des Matériaux Avancés (LMA), CNRS/IN2P3, F-69622 Villeurbanne, France
[66]Université Claude Bernard Lyon 1, F-69622 Villeurbanne, France

[67]Università di Napoli “Federico II”, Complesso Universitario di Monte S.Angelo, I-80126 Napoli, Italy
[68]NASA/Goddard Space Flight Center, Greenbelt, MD 20771, USA
[69]RESCEU, University of Tokyo, Tokyo, 113-0033, Japan
[70]Tsinghua University, Beijing 100084, China
[71]Texas Tech University, Lubbock, TX 79409, USA
[72]The Pennsylvania State University, University Park, PA 16802, USA
[73]National Tsing Hua University, Hsinchu City 30013, Taiwan, Republic of China
[74]Charles Sturt University, Wagga Wagga, New South Wales 2678, Australia
[75]University of Chicago, Chicago, IL 60637, USA
[76]Caltech CaRT, Pasadena, CA 91125, USA
[77]Korea Institute of Science and Technology Information, Daejeon 305-806, Korea
[78]Università di Roma “La Sapienza”, I-00185 Roma, Italy
[79]University of Brussels, Brussels 1050, Belgium
[80]Sonoma State University, Rohnert Park, CA 94928, USA
[81]Center for Interdisciplinary Exploration & Research in Astrophysics (CIERA), Northwestern University, Evanston, IL 60208, USA
[82]University of Minnesota, Minneapolis, MN 55455, USA
[83]The University of Melbourne, Parkville, Victoria 3010, Australia
[84]Institute for Plasma Research, Bhat, Gandhinagar 382428, India
[85]The University of Sheffield, Sheffield S10 2TN, United Kingdom
[86]West Virginia University, Morgantown, WV 26506, USA
[87]The University of Texas Rio Grande Valley, Brownsville, TX 78520, USA
[88]Università di Trento, Dipartimento di Fisica, I-38123 Povo, Trento, Italy
[89]INFN, Trento Institute for Fundamental Physics and Applications, I-38123 Povo, Trento, Italy
[90]Cardiff University, Cardiff CF24 3AA, United Kingdom
[91]Montclair State University, Montclair, NJ 07043, USA
[92]MTA Eötvös University, “Lendulet” Astrophysics Research Group, Budapest 1117, Hungary
[93]National Astronomical Observatory of Japan, 2-21-1 Osawa, Mitaka, Tokyo 181-8588, Japan
[94]School of Mathematics, University of Edinburgh, Edinburgh EH9 3FD, United Kingdom
[95]Indian Institute of Technology, Gandhinagar Ahmedabad, Gujarat 382424, India
[96]University of Szeged, Dóm tér 9, Szeged 6720, Hungary
[97]Embry-Riddle Aeronautical University, Prescott, AZ 86301, USA
[98]Tata Institute of Fundamental Research, Mumbai 400005, India
[99]INAF, Osservatorio Astronomico di Capodimonte, I-80131 Napoli, Italy
[100]University of Michigan, Ann Arbor, MI 48109, USA
[101]Rochester Institute of Technology, Rochester, NY 14623, USA
[102]NCSA, University of Illinois at Urbana-Champaign, Urbana, Illinois 61801, USA
[103]Universitat de les Illes Balears, IAC3—IEEC, E-07122 Palma de Mallorca, Spain
[104]University of Białystok, 15-424, Białystok, Poland
[105]SUPA, University of Strathclyde, Glasgow G1 1XQ, United Kingdom
[106]IISER-TVM, CET Campus, Trivandrum, Kerala 695016, India
[107]Canadian Institute for Theoretical Astrophysics, University of Toronto, Toronto, Ontario M5S 3H8, Canada
[108]Institute of Applied Physics, Nizhny Novgorod, 603950, Russia
[109]Pusan National University, Busan 609-735, Korea
[110]Hanyang University, Seoul 133-791, Korea
[111]University of Adelaide, Adelaide, South Australia 5005, Australia
[112]NCBJ, 05-400 Świerk-Otwock, Poland
[113]IM-PAN, 00-956 Warsaw, Poland
[114]Monash University, Victoria 3800, Australia
[115]Seoul National University, Seoul 151-742, Korea
[116]The Chinese University of Hong Kong, Shatin, NT, Hong Kong SAR, China
[117]University of Alabama in Huntsville, Huntsville, AL 35899, USA
[118]University of Massachusetts-Amherst, Amherst, MA 01003, USA
[119]ESPCI, CNRS, F-75005 Paris, France
[120]Università di Camerino, Dipartimento di Fisica, I-62032 Camerino, Italy
[121]Southern University and A&M College, Baton Rouge, LA 70813, USA
[122]College of William and Mary, Williamsburg, VA 23187, USA
[123]Montana State University, Bozeman, MT 59717, USA
[124]Instituto de Fí sica Teórica, University Estadual Paulista/ICTP South American Institute for Fundamental Research, São Paulo, SP 01140-070, Brazil
[125]University of Cambridge, Cambridge, CB2 1TN, United Kingdom
[126]IISER-Kolkata, Mohanpur, West Bengal 741252, India
[127]Rutherford Appleton Laboratory, HSIC, Chilton, Didcot, Oxon OX11 0QX, United Kingdom
[128]Whitman College, 345 Boyer Avenue, Walla Walla, WA 99362, USA
[129]National Institute for Mathematical Sciences, Daejeon 305-390, Korea
[130]Université de Lyon, F-69361 Lyon, France
[131]Hobart and William Smith Colleges, Geneva, NY 14456, USA
[132]Janusz Gil Institute of Astronomy, University of Zielona Góra, 65-265 Zielona, Góra, Poland
[133]King’s College London, University of London, London WC2R 2LS, United Kingdom
[134]Andrews University, Berrien Springs, MI 49104, USA
[135]Università di Siena, I-53100 Siena, Italy
[136]Trinity University, San Antonio, TX 78212, USA
[137]University of Washington, Seattle, WA 98195, USA
[138]Kenyon College, Gambier, OH 43022, USA
[139]Abilene Christian University, Abilene, TX 79699, USA